\documentclass[twocolumn,superscriptaddress,showpacs,showkeys,floatfix]{revtex4-2}
\usepackage{placeins}
\usepackage{graphicx}
\usepackage{tikz}
\usepackage{epstopdf}
\usepackage{mathtools}
\usepackage{dcolumn}
\usepackage{bm}
\usepackage{amsmath}
\usepackage{mathrsfs}

\usepackage{listings}%

\usepackage{xr-hyper}
\usepackage{xcite}
\externalcitedocument{SI_FePSe3_HP}
\usepackage[colorlinks=true,linkcolor=blue,citecolor=blue,urlcolor=blue]{hyperref}

\usepackage{units}

\usepackage{booktabs}
\usepackage{multirow}
\usepackage{soul}
\usepackage[normalem]{ulem}
\usepackage{makecell}
\usepackage{array}

\usepackage[utf8]{inputenc}   
\renewcommand{\selectlanguage}[1]{}

\usepackage{xcolor} 
\ifx\nocolor\undefined 
    
   \newcommand{\tred}[1]{\textcolor{black}{#1}} 
   
   \newcommand{\sdelete}[1]{} 

\else
   
   \newcommand{\tred}[1]{{#1}}
   
\fi

\newcommand{\TMPX}{$TM$P$X_3$}	

\makeatletter
\renewcommand\paragraph{\@startsection{paragraph}{4}{\z@}%
  {2.5ex \@plus 1ex \@minus .2ex}%
  {-1em}%
  {\normalfont\normalsize\bfseries}}  

\makeatother

\newcommand{\replyv}[1]{\textcolor{black}{#1}}
\newcommand{\replyvs}[1]{}

\begin{document}

\title{Emergent Polar Metal Phase in a Van der Waals Mott Magnet}

\author{Shiyu Deng}
\altaffiliation{These authors contributed equally.}
\email{Email: sd864@cantab.ac.uk}
\affiliation{Cavendish Laboratory, University of Cambridge,  Cambridge, CB3 0HE, United Kingdom}
\affiliation{Institut Laue-Langevin, 71 Avenue des Martyrs, 38000 Grenoble, France}

\author{Matthew J. Coak}
\altaffiliation{These authors contributed equally.}
\email{Email: m.j.coak@bham.ac.uk}
\affiliation{School of Physics and Astronomy, University of Birmingham, Birmingham B15 2TT, United Kingdom}

\author{Charles R. S. Haines}
\affiliation{School of Engineering, Mathematics and Physics, University of East Anglia, Norwich Research Park, Norwich, Norfolk, NR4 7TJ, United Kingdom}

\author{Hayrullo Hamidov}
\affiliation{Cavendish Laboratory, University of Cambridge, Cambridge, CB3 0HE, United Kingdom}
\affiliation{Navoi State University of Mining and Technologies, \replyv{76V Galaba Avenue}\replyvs{72 M. Tarobiy Street}, Navoi 210100, Uzbekistan}

\author{Giulio I. Lampronti}
\affiliation{Department of Earth Sciences, University of Cambridge, Cambridge CB2 3EQ, United Kingdom}

\author{David M. Jarvis}
\affiliation{Cavendish Laboratory, University of Cambridge, Cambridge, CB3 0HE, United Kingdom}
\affiliation{Institut Laue-Langevin, 71 Avenue des Martyrs, 38000 Grenoble, France}

\author{Xiaotian Zhang}
\affiliation{Cavendish Laboratory, University of Cambridge, Cambridge, CB3 0HE, United Kingdom}

\author{Cheng Liu}
\affiliation{Cavendish Laboratory, University of Cambridge, Cambridge, CB3 0HE, United Kingdom}

\author{Dominik Daisenberger}
\affiliation{Diamond Light Source, Harwell Science and Innovation Campus, Didcot, OX11 0DE, United Kingdom}

\author{Mark R. Warren}
\affiliation{Diamond Light Source, Harwell Science and Innovation Campus, Didcot, OX11 0DE, United Kingdom}

\author{Thomas C Hansen}
\affiliation{Institut Laue-Langevin, 71 Avenue des Martyrs, 38000 Grenoble, France}

\author{Stefan Klotz}
\affiliation{Sorbonne Universite, IMPMC, CNRS, UMR 7590, 4 Place Jussieu, 75252 Paris, France}

\author{Chaebin Kim}
\affiliation{Department of Physics and Astronomy, Seoul National University, Seoul 08826, Republic of Korea}

\author{Pengtao Yang}
\affiliation{Beijing National Laboratory for Condensed Matter Physics and Institute of Physics, Chinese Academy of Sciences, Beijing 100190, China}
\affiliation{School of Physical Sciences, University of Chinese Academy of Sciences, Beijing 100190, China}

\author{Bosen Wang}
\affiliation{Beijing National Laboratory for Condensed Matter Physics and Institute of Physics, Chinese Academy of Sciences, Beijing 100190, China}
\affiliation{School of Physical Sciences, University of Chinese Academy of Sciences, Beijing 100190, China}

\author{Jinguang Cheng}
\affiliation{Beijing National Laboratory for Condensed Matter Physics and Institute of Physics, Chinese Academy of Sciences, Beijing 100190, China}
\affiliation{School of Physical Sciences, University of Chinese Academy of Sciences, Beijing 100190, China}

\author{Je-Geun Park}
\affiliation{Department of Physics and Astronomy, Seoul National University, Seoul 08826, Republic of Korea}

\author{Andrew R. Wildes}
\affiliation{Institut Laue-Langevin, 71 Avenue des Martyrs, 38000 Grenoble, France}

\author{Siddharth S Saxena}
\email{Email: sss21@cam.ac.uk}
\affiliation{Cavendish Laboratory, University of Cambridge, Cambridge, CB3 0HE, United Kingdom}
\affiliation{British Management University Tashkent, 35 Mirza Bobur Street, Tashkent, Uzbekistan}
\affiliation{\replyv{Kazakh-British Technical University, Almaty 050000, Kazakhstan}}

\date{\today}

\begin{abstract}
\section*{Abstract}
We report the emergence of a two-dimensional
\replyvs{(2D)}\replyvs{polar} metal phase \replyv{with a polar space
  group} in van der Waals compound FePSe$_3$, \replyvs{under moderate
  pressures.  This layered material is} a\replyv{n antiferromagnetic}
Mott insulator \replyvs{with antiferromagnetic order} under ambient
conditions.  We show that FePSe$_3$ \replyvs{uniquely allows tuning}
\replyv{can be tuned from} a \replyvs{2D} correlated insulator into an
exotic metal, where a loss of inversion symmetry leads to periodic
polar displacements of ions, within a conducting phase - a polar
metal.  Our combined synchrotron and neutron diffraction data
\replyvs{allow us to }present \replyvs{a long-sought,} an unambiguous
high-pressure structural model\replyv{, where the system switches from
  being centrosymmetric into a polar space group.} \replyvs{and show
  the polar displacements of this new phase.  We also observe the
  suppression of magnetic ordering at the insulator-to-metal
  transition correspondent with this structural change.  Our work
  outlines a comprehensive temperature-pressure phase diagram of
  FePSe$_3$, combining detailed structural, magnetic and transport
  data.}  The high-pressure phase exhibits \replyvs{activated
  semiconductor behavior at high temperatures,} a $T^2$-dependence
\replyv{of}\replyvs{in its} resistivity at lower temperatures --
despite the \replyv{lack of ideal} conditions \replyvs{required} for a
\replyvs{`good metal'} Fermi-Liquid description \replyv{to be
  applicable} \replyvs{not being met in this case} -- and a
low-temperature resistivity upturn which is suppressed as the system
is tuned away from the concomitant transitions. \replyvs{In this work,
  the realisation of a tunable 2D polar metal state in FePSe$_3$ due
  to the loss of its inversion symmetry combined with pressure-induced
  metallicity} \replyv{This work} offers a promising \replyvs{new} platform to
investigate this exotic phase \replyv{in low dimensionality} at
accessible pressures.
\end{abstract}

\maketitle
\section{Introduction}

The study of layered van der Waals (vdW) compounds has become
increasingly popular since the revolutionary discovery of
graphene~\cite{2005_Zhang_Graphene}.  Magnetic vdW materials bring yet
more functionality and fundamental physics to explore.  In their
atomically thin versions, they have attracted particular attention for
applications in spintronics and
magneto-optics~\cite{2016_MagneticGraphene_outlook,
  2022_MagGenome_2DvdW}.  They also offer an excellent platform for
studying strongly correlated electrons in low dimensions.

Many vdW magnets are also insulating, but undergo an
insulator-to-metal transition (IMT) under the application of
pressure~\cite{2018_Haines_highPressure_XRD_FePS3,2022_MagGenome_2DvdW}.
This makes them conceptually similar to the parent compound of
cuprates: being insulating two-dimensional (2D) antiferromagnets
\tred{with strongly correlated electrons}.  In cuprates, varying
stoichiometry drives an insulator to a `strange metal', and eventually
superconducting phase~\cite{1986_First_Cuprate,1988_Cuprate_120K}.
Varying the pressure applied to layered vdW compounds offers a cleaner
and more controllable way to tune physical properties, providing great
insight into the underlying physics of the IMT and any exotic phases
that may emerge
~\cite{2020_Corentin_SC_polar_modes_FE,2021_Pressure_FE_tran_Polar_Metal,2023_PRM_MJames_PolarMetals}.

\replyv{Among those exotic phases},
polar metals have become the focus of significant recent work due to their
counterintuitive combination of both metallic conductivity and a
non-centrosymmetric polar structure giving an ordered polarization
(electric dipoles).  Despite predictions in the
60s~\cite{Anderson_PolarMetals_1965}, only in the last decade or so
have experimental realizations of polar metals been confirmed.  There
are few~\cite{bhowal_polar_2023}, if any, examples of such systems
being tunable via a ‘clean’ parameter such as applied hydrostatic
pressure\replyv{~\cite{2023_PRM_MJames_PolarMetals,Jager2024}}.

One intriguing \replyv{strongly correlated vdW magnet} family is the
transition metal chalcogenophosphates {\TMPX} ($TM$ = Mn, Fe, Co, Ni,
$etc.$, $X$ = S, Se).  These insulating compounds are antiferromagnets
at ambient pressure, with the magnetic $TM^{2+}$ ions forming a
honeycomb structure within the vdW
layers~\cite{klingen1973_FePX3_structure,1979_Brec_TMPX3_PhysProperties,1981_Wiedenmann_neutron_MnPSe3_FePSe3}.
Previous studies have revealed pressure-induced transitions from
insulating to metallic, and, in some reports, superconducting
states~\cite{2018_Haines_highPressure_XRD_FePS3,
  2018_Wang_pressure_SC_FePX3,
  2020_Coak_TuneDimensionality_TMPS3,2023_NiPSe3_Pressure_SC,2023_SnPS3_pressure_FE2SC,2024_SnPSe3_pressure_SC,2024_CrSbSe3_pressure_SC}.
However, the critical underlying structural transitions remain
elusive, prohibiting further investigation and engineering of this
compound family. Up to this point, the literature has contained
contradictory results and predictions - care must be taken in
interpreting observed data, as demonstrated in this work and earlier
studies~\cite{2023_Jarvis_comparative_single_powder_FePS3,2024_MPX3_Pressure_review_Mandrus}

We selected FePSe$_3$ as the focus of this work, which was recently
reported to exhibit superconductivity at approximately 2.5~K and
9.0~GPa~\cite{2018_Wang_pressure_SC_FePX3}. At ambient pressure, the
crystal structure of FePSe$_3$ consists of two layers of Se on a
triangular lattice, sandwiching one layer of Fe on a honeycomb
lattice, with P$_2$ dimers located at the centre of each
Fe-hexagon~\cite{klingen1973_FePX3_structure}.  Both P$_2$ dimers and
Fe ions are octahedrally coordinated with neighboring Se.
The layers stack in an $ABCABC$ sequence, with Fe and P$_2$ aligned
along axes parallel to the $c$-axis, which is normal to the vdW
layers.  FePSe$_3$ crystallizes in a centrosymmetric rhombohedral
space group $R\overline{3}$.  Below the reported $T_N$ of $108\sim 119$~K,
it orders antiferromagnetically with a propagation vector of
($\nicefrac{1}{2}$,0,\replyvs{$\pm$}$\nicefrac{1}{2}$) and spin
moments aligned parallel to the
$c$-axis~\cite{1981_Wiedenmann_neutron_MnPSe3_FePSe3,
  2023_FePSe3_PhononSymm_Spin-Phonon-Coupling}.

\begin{figure}   
    \centering
    \includegraphics[width=0.49\textwidth]{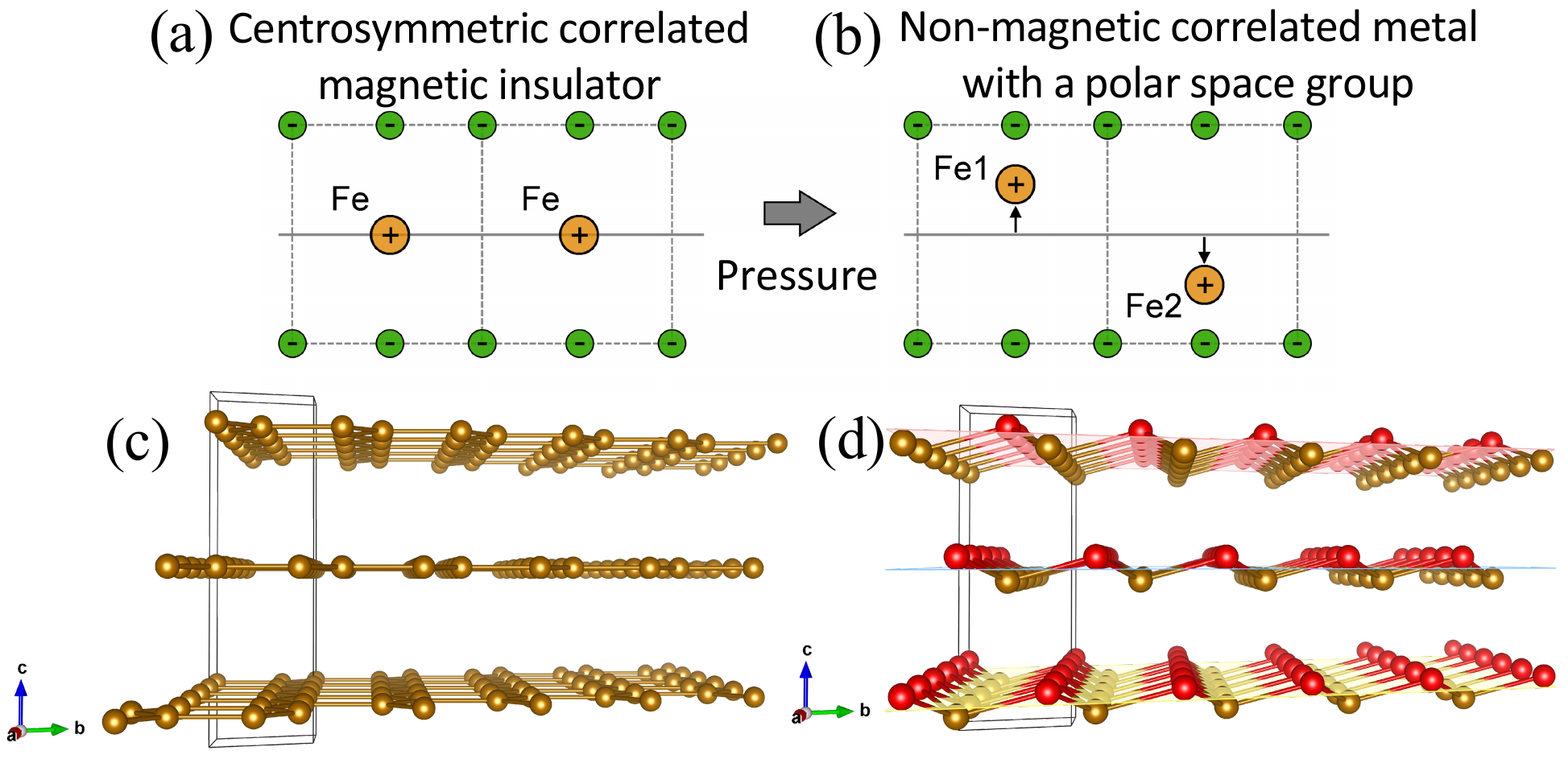}   
    \caption{ \textbf{Tuning a Centrosymmetric Mott Insulator to a
        Correlated \replyvs{Polar} Metal \replyv{with a Polar Space
          Group}} This illustration demonstrates our key findings:
      tuning of a 2D material from a magnetic insulator (a,c) to a
      non-magnetic \replyvs{polar} metal \replyv{with a polar space
        group} (b,d) via pressure. The latter is a special state
      that not only exhibits ordered displacements of polar ions, but
      also free carriers which rearrange so as to screen long-range
      electrical polarisation. \replyv{In (a,b), Fe atoms are
        shown as brown balls with positive (+) charges while Se atoms
        are shown as green balls with negative (-) charges.  
        In (c,d), we only present Fe atoms for clarity.  Under
        pressure, the Fe atoms split into two distinct Wyckoff sites,
        as highlighted in brown and red respectively in (d).}}
    \label{Fig1-Cartoon}
\end{figure}

In this work, we present a multi-probe study of FePSe$_3$, combining
synchrotron and neutron diffraction with resistivity measurements to
investigate the effect of hydrostatic pressure on the crystalline
structure, magnetic ordering, and electronic transport properties.  We
combine our results into a detailed phase diagram and report direct
evidence for the emergence of a \replyvs{new polar
  metal}\replyv{high-pressure} phase\replyv{, where metallicity is}
linked to the loss of inversion symmetry\replyvs{ in the high-pressure
  structure}\replyv{~(see Fig.~\ref{Fig1-Cartoon}), suggesting a polar
  metal}.
We also discuss the loss of magnetic order as this state
is entered, and the puzzles in the resistivity found in transport
measurements.

\section{Results}
\begin{figure*}[htbp]
    \centering
    \includegraphics[width=0.99\textwidth,angle=0]{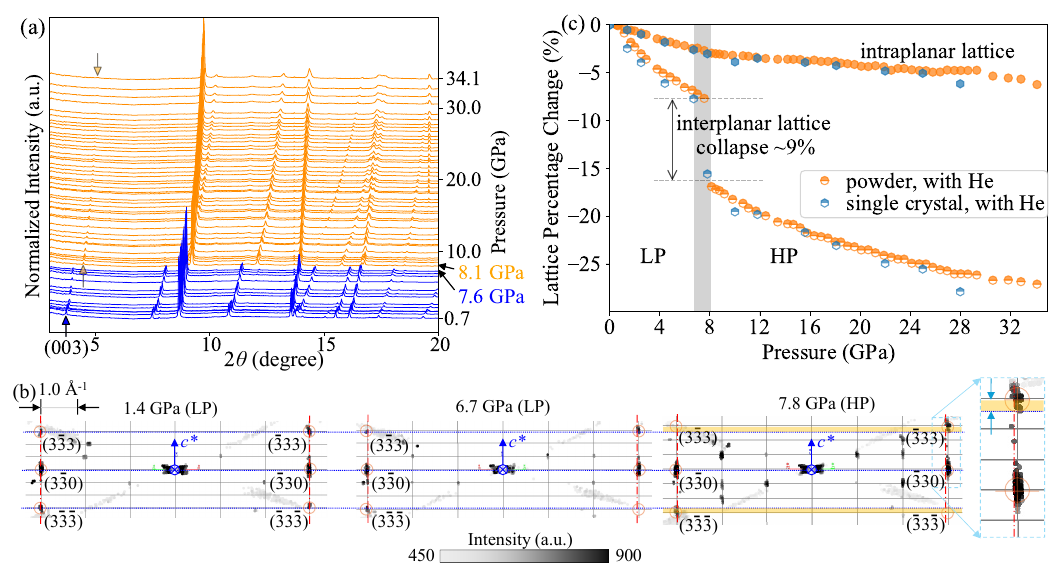}
    \caption{ \textbf{Powder and Single-Crystal Synchrotron
        Diffraction Results and Lattice Parameters Evolution under
        Pressure} (a) Integrated diffraction intensities as a function
      of $2\theta$ for FePSe$_3$ powder sample in a DAC loaded with
      helium gas as a pressure-transmitting medium. For each pressure,
      the intensity was normalized to its maximum, and then an offset
      proportional to the pressure value was added. Blue and orange
      represent data assigned to the LP and HP phases, respectively.
      (b) Slices \replyv{through the single-crystal data}\replyvs{of }
      ($h\overline{h}l$) planes at the lowest pressure (\unit[1.4]{GPa}, LP),
      before (\unit[6.7]{GPa}, LP) and after (\unit[7.8]{GPa}, LP) the phase
      transition pressure. The sub-pictures are all scaled in
      {\AA}$^{-1}$. The red and blue lines are given to guide the eye
      to show the separation distance at the \unit[1.4]{GPa}, LP phase. The
      yellow color blocks highlight the difference in the separation
      of $c^*$. A zoom-in view is provided at \unit[7.8]{GPa} to better
      illustrate the abrupt changes in $c^*$.  (c) Pressure-induced
      relative changes of the lattice parameters in reference to the
      ambient-pressure structure. Orange and blue represent powder and
      single crystal cases, respectively. \replyv{The fully filled
        dots represent the intraplanar lattice spacings while the
        half-filled ones represent the interplanar.} The grey-shaded
      region indicates the interplanar lattice collapse transition
      pressure.}
    \label{Fig_FePSe3_Xray_LatticeChange}
\end{figure*}

In this section we first present our powder and single-crystal X-ray
diffraction data with varying pressure, \replyvs{to unambiguously
  demonstrate a $c$ axis collapse at the IMT after controversy in
  the literature and }combining datasets to solve the high-pressure
structure \replyv{and to demonstrate a $c$ axis collapse at the IMT
  after controversy in the literature}. We then show through
neutron diffraction that magnetic order is lost entirely in this
phase, and finally present electrical transport data across the phase
diagram. \replyvs{In the Discussion we build on this range of observations to
demonstrate discovery of \replyv{this previously undiscovered}
\replyvs{the new polar metal} phase and the various puzzles uncovered
in its transport data.}

\subsection{Synchrotron Diffraction and Crystalline Structures}

\subsubsection{Powder Diffraction} 
The powder diffraction rings of FePSe$_3$ loaded in helium gas as the
\replyv{pressure-transmitting medium}
(PTM) showed no significant intensity variation about the Debye-Scherrer
cone, indicating that our sample loading procedure was free from
strong preferred orientation.  The details and subsequent data
reductions converting the original diffraction rings into integrated
diffraction intensities as a function of 2$\theta$ are given in the
\replyv{Supplementary Fig.~\ref{SI:fig_powder_withHe_DiffRing}}.

Fig.~\ref{Fig_FePSe3_Xray_LatticeChange}~(a) presents an evolution of
the post-processed powder diffraction patterns across a range of
pressures from 0.7 to 34.1~GPa. The integrated intensities are
normalized to the maximum intensity for each pressure point. All
datasets are shifted vertically by an offset proportional to the
pressure value.

The data show a dramatic shift in the 2$\theta$ position of a number
of peaks between 7.6 and 8.1~GPa which indicates a distinct phase
transition. The transition is sharp with little evidence of phase
co-existence. There is no peak broadening associated with the
transition, suggesting that a single-phase model with minimal
distortion may be used to determine the crystal structure of the
high-pressure phase. No further phase transition was evident up to the
highest measured pressure~(\unit[34.1]{GPa}). The low- and high-pressure
phases will subsequently be referred to as LP and HP, respectively.
Data assigned to these two phases are plotted in blue and orange in
Fig.~\ref{Fig_FePSe3_Xray_LatticeChange}~(a) for a straightforward
comparison.

The LP Bragg peak at the lowest $2\theta$ is indexed in the refinement
as 003, as marked in Fig.~\ref{Fig_FePSe3_Xray_LatticeChange}~(a).  It
is reasonable to assume that the crystal structure does not undergo a
substantial reconstruction during the transition from LP to
HP. Consequently, the position of this peak can be assumed to be due
to the change of the interplanar distance at all pressures. The peak
undergoes a drastic change at the phase transition, indicating that
the interplanar spacing collapses at the LP-HP transition.

The LP data were analyzed using the ambient pressure structure model,
which had the hexagonal $R\overline{3}$ space
group~\cite{1979_Brec_TMPX3_PhysProperties}.  The refined lattice
parameters at the lowest pressure point (\unit[0.7]{GPa}) were
determined to be $a = b =$ \unit[6.238(2)]{\AA} and $c =$
\unit[19.49(1)]{\AA}, which are in reasonable agreement with the
literature values at ambient pressure ($a = b =$ \unit[6.265(6)]{\AA},
$c =$ \unit[19.80(2)]{\AA}, and $\gamma =$ \unit[120]{$^{\circ}$}).
The HP data were analyzed using the structural model we obtained from
our single-crystal synchrotron studies.

\subsubsection{Single-Crystal Diffraction} 
In the single-crystal experiment, we collected high-quality
diffraction data from 1.4 to \unit[28.0]{GPa}. Slices of the
($h\overline{h}l$) scattering plane exhibit strong-intensity peaks
which are indexed as $3\overline{3}l$ and $\overline{3}3l$ where $l$ =
0, $\pm3$, as shown in
Fig.~\ref{Fig_FePSe3_Xray_LatticeChange}(b). The slices are plotted
to scale in {\AA}$^{-1}$.  The intensity of the peaks is consistent
with the structure factor analysis.  The separation between reflection
index points peaks along the {[001]} and {[1$\overline{1}$0]}
directions unambiguously reflects the changes of the inter- and
intraplanar lattice parameters, respectively. An abrupt increase in
separation along the {[001]} direction is clearly seen from 6.7 to
\unit[7.8]{GPa} (highlighted with yellow color blocks), while changes
along the {[1$\overline{1}$0]} direction remain negligible throughout
the entire pressure range.  This observation in the raw data,
independent of any structural model, provides definitive and
model-independent proof of the interplanar collapse with applied
hydrostatic pressure in FePSe$_3$.

The lattice parameters were initially refined based on indexed
reflections. We used the ambient-pressure lattice parameters of LP as
the starting point to refine the cell at the lowest pressure
(\unit[1.4]{GPa}). No space group symmetry was forced on the
refinement at this step. The well-indexed reflection peaks and
refined lattice parameters (\unit[{\it a}=6.185(4)]{\AA},
\unit[{\it b}=6.176(3)]{\AA}, \unit[{\it c}=18.89(3)]{\AA},
\unit[$\alpha$=90.28(1)]{$^\circ$}, \unit[$\beta$=89.90(1)]{$^\circ$}
and \unit[$\gamma$=119.99(7)]{$^\circ$}) are consistent with the
$R\overline{3}$ symmetry of LP with tolerable deviations. We
consistently refined the lattice parameters for all pressures using
the cell parameters at the previous pressure point. The results are
summarized in \replyv{Supplementary
  Table}~\ref{SI:table_FePSe3_single_crystal}.

Fig.~\ref{Fig_FePSe3_Xray_LatticeChange}(c) summarizes the relative
changes of lattice parameters as a function of pressure, resulting
from our comprehensive synchrotron studies.  It can be seen that the
intraplanar lattice parameters decrease smoothly at an average rate of
{$\sim0.17\%$ per GPa}.  The interplanar distance, by contrast,
declines more rapidly at an average rate of {$\sim0.86\%$ per GPa},
followed by an abrupt drop of approximately 9\% across the LP-HP
transition.  After the transition, {\it c} continues to decrease at a rate
of {$\sim0.50\%$ per GPa} up to the highest pressure. The LP-HP
transition region is shaded to guide the eye.

Thorough analysis of the single-crystal diffraction is the most
reliable way to assign the correct space group and determine the
structural model.  Data below \unit[6.7]{GPa} were well-refined with
the LP structure model of the $R\overline{3}$ space group.
Refinements of the data above \unit[7.8]{GPa} with the $R\overline{3}$
space group gave poor quality parameters. As can be seen from
Fig~\ref{Fig_FePSe3_Xray_LatticeChange}(b), there is no substantial
change in {\it a} and {\it b}, and merely a sudden change of
interplanar spacing in {\it c}.  Hence it is sensible to keep using
the hexagonal cell description and check among the subgroups of
$R\overline{3}$ with {\it k}-index=1, namely $R3$, $P\overline{1}$ and
$P1$.
After careful inspection, we find that there are no new peaks at new
$d$ spacings, but peaks with equivalent $d$ spacing present different
structure factors.  Further analysis determined that this is due to a
Friedel Pair\replyv{~\cite{friedel1913}} not showing the same observed
structure factor, namely $I(h k l)\neq I(\bar{h} \bar{k}
\bar{l})$\replyv{~\cite{bijvoet1951}}. \replyv{ Quantitative analysis
  shows that the LP phase (\unit[4.4]{GPa}) yields a $Z$-distribution
  consistent with a Gaussian noise centered at $Z=0$, while the HP phase
  (\unit[10.0]{GPa}) exhibits a signifiant deviation, as shown in
  Supplementary Fig.~\ref{SI:FriedelPair}.} This can only be explained
by the loss of centrosymmetric symmetry, leaving $R3$ and $P1$ as the
only possible solutions.
Meanwhile, we still only observe intensities at $-h + k + l = 3n$,
suggesting that the hexagonal setting remains. \replyvs{Consequently,
  we demonstrate the sole possible solution for the HP phase to be the
  $R3$ space group.} All peaks can be indexed \replyv{with a
  rhombohedral unit cell and their intensities are compatible}
\replyvs{and refined} with the $R3$ space group\replyv{, which has
  been identified as the correct space group}. \replyv{For
  completeness, we show the observed versus calculated structure
  factors graphically in Supplementary
  Fig.~\ref{SI:Fig_HP_StructureFactor}, and the refinement parameters
  in Supplementary Table~\ref{SI:Table_10GPa_compare}.}

Two representative sets of crystallographic information for the LP and
HP phases, including refinement parameters, are given in
Table~\ref{Table_LPHP_CrystalInfo}.  The \replyv{resulting} atomic
position information \replyv{is listed in Supplementary
  Table~\ref{SI:Tab_FePSe3_WyckoffPos_LP} and
  Table~\ref{SI:Tab_FePSe3_WyckoffPos_HP}.}\replyvs{, observed versus
  calculated structure factors, as well as the analysis of Friedel
  pairs graphically are given in the SI~\cite{Ref_SM}.}

The solved $R3$ model of the HP phase can be used equally successfully
to refine powder data, providing \replyv{further confirmation of the
  correctness of the structural model and} \replyvs{an} unambiguous
\replyv{lattice parameters of the} HP phase \replyvs{structure model}
\replyv{over the pressure range investigated}.

\begin{table}[htbp]
\centering
\caption{ \textbf{Reduced crystallographic data in LP (\unit[4.4]{GPa}) and HP
    (10.0, \unit[28.0]{GPa}) phases.}  The parameters are obtained from the
  single-crystal synchrotron diffraction experiment.  Here we use the primitive
  hexagon cell description for the lattice parameters.}
\label{Table_LPHP_CrystalInfo}
   \begin{tabular}{lcccc}  
   \hline \hline \rule{0pt}{2.5ex}{\hfill}
                 &    & \unit[4.4]{GPa} & \unit[10.0]{GPa}  &  \unit[28.0]{GPa}\\ 
   \midrule
   \multirow{3}{1.8cm}{Lattice parameters}
   &Space group       & $R\overline{3}$ & $R3$ & $R3$\\
   &$a=b$ ({\AA})     & 6.0971(6)       & 6.0155(7) & 	5.895(3)\\
   &$c$ ({\AA})       & 18.279(19)      & 15.34(6)  & 	14.1(2)\\
   &Volume ({\AA}$^3$)& 588.5 (6)       & 480.8 (18) & 423(7)\\ 
   \midrule
   \multirow{4}{1.8cm}{Refinement parameters}   
   &\textbf{$R[F^2 > 2{\sigma}(F^2)]$} & 0.0038 & 0.090 &0.0922\\
   &\textbf{$wR(F^2)$}                 & 0.095  & 0.221 & 0.2401\\
   &\textbf{$S$}                       & 0.88   & 2.1 & 2.3\\
   &Reflections                        & 36     & 125 & 134 \\
   \hline \hline 
   \end{tabular}
\end{table}

\begin{figure*}[htbp]
    \centering
    \includegraphics[width=0.9\textwidth,angle=0]{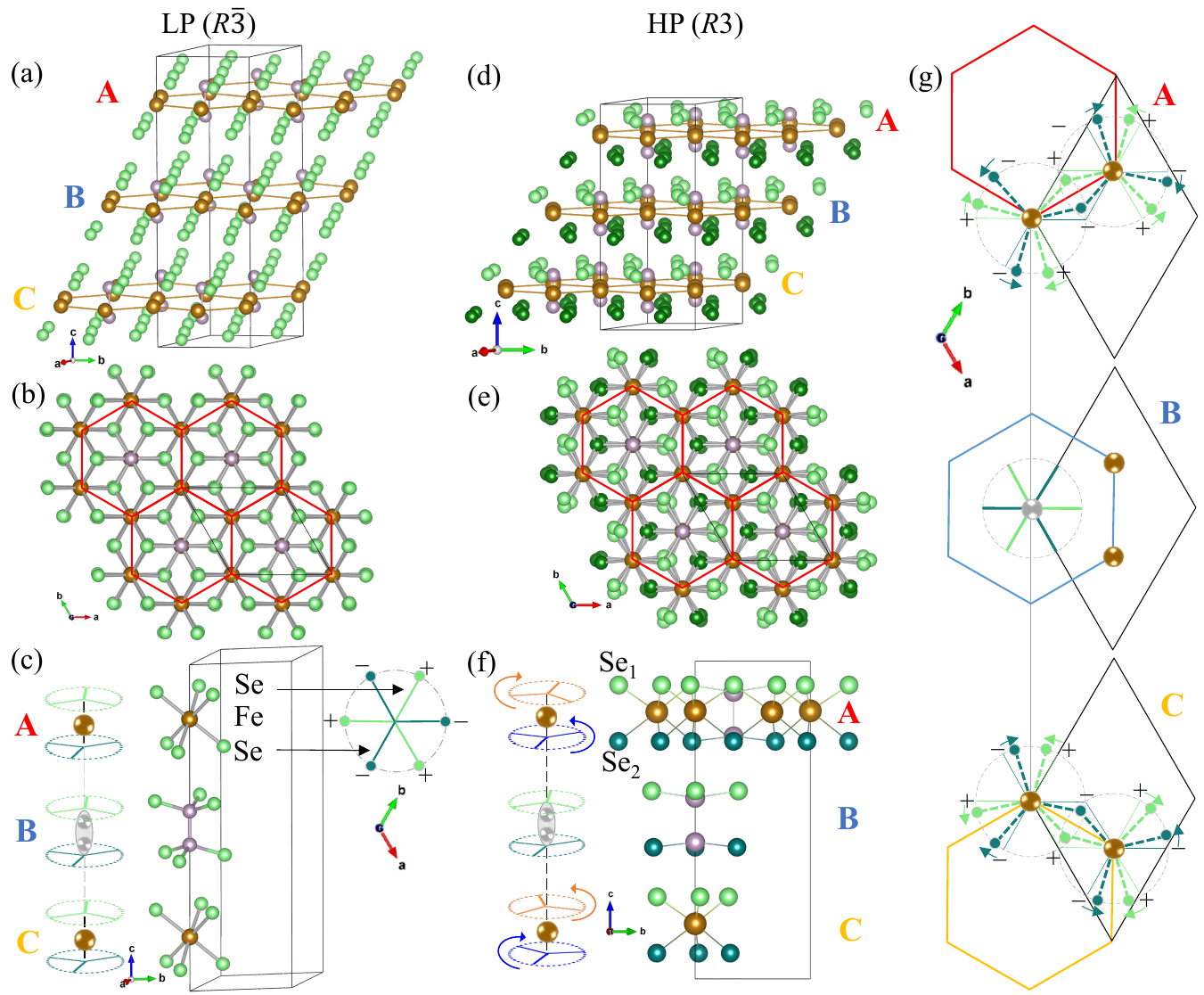}
    \caption{
      \textbf{Crystalline structures of the
        ambient-/low-pressure (LP) and high-pressure (HP) phases of
        FePSe$_3$} Fe: brown; P: grey; Se: green. (a,
      \replyv{d}\replyvs{b}) are the structural isometric view with a
      minimal stacking unit of $ABC$ triple layers; (b, e) display the
      view normal to the vdW planes, with the red solid lines
      indicating the Fe-honeycomb network; (c, f) present the FeSe$_6$
      and P$_2$Se$_6$ octahedra along the axis parallel to
      \textit{c}. The cartoon aside depicts Se planes in both phases.
      The light green and dark green with + and - signs represent the
      Se atoms in the plane above or below the plane defined by Fe or
      P$_2$ dimer within each vdW layer.  The collective displacement
      of Se atoms in the HP phase is demonstrated with different
      colors and arrows.  (g) shows the normal view of the three
      individual stacking layers in $ABC$ ordering. The collective
      rotation of each Se plane is indicated with arrows in different
      colors.}
\label{Fig_FePSe3_StructureDemo}
\end{figure*}

\subsubsection{Key Structural Differences}
Fig.~\ref{Fig_FePSe3_StructureDemo} exhibits the crystalline structure
of LP and HP FePSe$_3$ side by side. (a) and (d) provide an isometric
view of the bulk structures. Fe, P and Se atoms are shown in brown,
gray and green respectively.  We represent the Fe-hexagon network with
solid brown lines.  The interplanar distances shrink substantially
from LP to HP, while the stacking geometry $ABCABC$ remains unchanged.
Normal to the vdW layers, as shown in (b) and (e), the Fe-hexagon
network remains undistorted from LP to HP.  The $C3$ rotation symmetry
is thus preserved across the transition.

The HP phase differs from the LP phase primarily in loss of the
inversion symmetry, resulting from the buckling of the Fe-hexagons
along the $c$-axis and the collective displacement of Se sites.  The
LP phase only possesses one Wyckoff site for Se (18$f$), thus forming
perfect octahedral coordination with Fe or $\rm P_2$ positioned in the
center, as is shown in Fig.~\ref{Fig_FePSe3_StructureDemo}(c).
Within each vdW layer, Fe atoms and the center of the P$_2$ dimers
define one plane at $z$ ($z=\nicefrac{1}{6}+\nicefrac{n}{3}$, n:
integral).  Se atoms at $z \pm \Delta$ positions form two sandwiching
planes, represented by light and dark green colors together with + and
- signs for $z+\Delta$ and $z-\Delta$, respectively.  We also show a
normal view to the $ab$ plane of one Se-octahedron on the top-right
corner of Fig.~\ref{Fig_FePSe3_StructureDemo}(c).  The bonds between
Fe or P$_2$ and the three Se atoms at the same plane ($z+\Delta$ or
$z-\Delta$) are shown by contrasting colors.

In the HP phase, we observe that the Fe planes buckle slightly,
resulting from Fe$_1$($3a$) moving upward while Fe$_2$($3a$) move
downward by $\sim1\%$ in lattice unit along the $c$-axis, in reference
to their center position.  The Se atoms also split into two distinct
Wyckoff sites, Se$_1$ (9$b$) forming one Se plane and Se$_2$ (9$b$)
forming the other, as highlighted in
Fig.~\ref{Fig_FePSe3_StructureDemo}(d-f) by light and dark green
colors respectively.  The relative displacement of the two Se planes
to the center of the Fe plane differs slightly.  Within each
[FeSe$_6$] octahedron, Se$_1$ and Se$_2$ are rotated coherently within
the plane but in opposite directions between the planes.  Meanwhile,
the [P$_2$Se$_6$] octahedron remains nearly undistorted.  We indicate
the local in-plane collective rotations of Se atoms with orange and
blue along one axis parallel to the $c$-axis in
Fig.~\ref{Fig_FePSe3_StructureDemo}(f).  In addition,
Fig.~\ref{Fig_FePSe3_StructureDemo}(g) exhibits how the octahedron's
local distortion propagates within and across the vdW planes in a bulk
structure.  We use red, blue and gold hexagon networks to present the
Fe-hexagon in $ABC$ stacking sequence.  It can be seen that each
neighboring Se-triangle rotates in the opposite direction within one
Se plane.  These structural distortions break the inversion symmetry
of the crystalline structure, leaving \replyv{the} FePSe$_3$ HP phase
in the $R3$ space group with a polar point group 3, and also
contribute to a net dipole moment along the $c$-axis.

\subsection{Neutron Scattering and Magnetic Properties}

\begin{figure}[htbp]  
\centering
    \includegraphics[width=0.49\textwidth,angle=0]{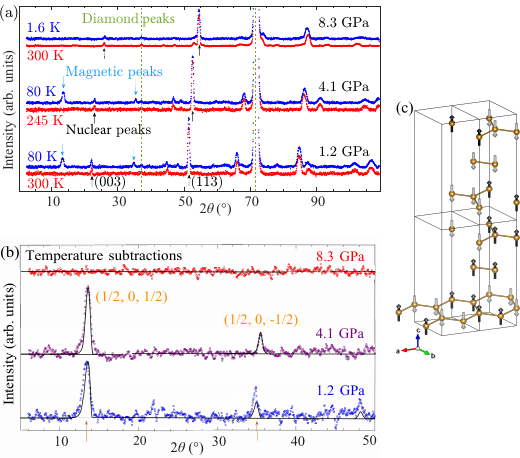}
    \caption{\textbf{Powder Neutron Diffraction Patterns and Magnetic
        Structure under Pressure} (a) Diffraction data taken at
      temperatures well above (red) and below (blue) the magnetic
      transition temperature at pressures of 1.2, 4.1 and
      \unit[8.3]{GPa} after background subtraction.  Nuclear, magnetic
      peaks for FePSe$_3$, and peaks originating from the diamond
      anvils are indicated with black, cyan and green, respectively.
      (b) Resulting diffraction intensity obtained by subtracting the
      intensity at high temperature (red) from that at low temperature
      (blue) for each pressure point.  Magnetic peaks
      \replyvs{consistent with the ambient-pressure}
      $(\nicefrac{1}{2},0,\pm\nicefrac{1}{2})$
      \replyvs{magnetic propagation vectors} are marked
      with orange arrows\replyv{, both associated and consistent with the 
      ambient-pressure propagation vector $(\nicefrac{1}{2},0,\nicefrac{1}{2})$}.
      Solid black lines represent refinements of
      the data. (c) Magnetic ordering of Fe atoms in the LP phase
      with the nuclear unit cell, in solid black lines as a reference.
      The black and white arrows on $\rm Fe^{2+}$ atoms show opposing
      spin moments, parallel to the $c$-axis. \tred{Not all magnetic
        atoms are shown in the magnetic cell, which are four times the
        nuclear cell.}  }
    \label{Fig_FePSe3_MagDiffraction_pressure}
\end{figure}

Fig.~\ref{Fig_FePSe3_MagDiffraction_pressure}(a) shows the powder
neutron diffraction patterns after background subtraction at
temperatures above (HT) and below (LT) the magnetic transition in red
and blue respectively, for pressure values of 1.2, 4.1 and
\unit[8.3]{GPa}.  In addition to the sample signals, there were
visible peaks originating from the diamond anvils, as marked with
green dotted lines in the graph.  More details about background
subtraction and diamond peak identification can be found in the
\replyv{Supplementary Section II.} \replyvs{SI~\cite{Ref_SM}}

The HT data (red) contained only the nuclear peaks. The indexed
crystal structures were consistent with our synchrotron results. The
pressure values for the neutron measurements were then determined
using the structure parameters resulting from the synchrotron data.
The LT data (blue) included both nuclear and magnetic peaks.
\replyv{We performed the full refinements including both nuclear and
  magnetic contributions.} Since the nuclear peaks remained nearly
unchanged with varying temperatures at each pressure point, we
subtracted the HT data from the LT data to separate the magnetic
peaks. The resulting temperature subtractions are shown in
Fig.~\ref{Fig_FePSe3_MagDiffraction_pressure}(b).

At 1.2 and \unit[4.1]{GPa}, pressures below the LP to HP crystalline
transition, magnetic peaks can be seen clearly and well-indexed with
the ambient-pressure magnetic structure model by Wiedenmann {\it et
  al.}~\cite{1981_Wiedenmann_neutron_MnPSe3_FePSe3} assuming no
substantial changes in magnetic structures.  The Bragg peaks
corresponding to the $(\nicefrac{1}{2},0,\nicefrac{1}{2})$
\replyvs{and $(\nicefrac{1}{2},0,\nicefrac{\overline{1}}{2})$}
antiferromagnetic propagation \textit{k}-vector\replyvs{s are}
\replyv{are} marked in
Fig.~\ref{Fig_FePSe3_MagDiffraction_pressure}~(b). \replyv{The
  magnetic structure was refined using the $P_S\overline{1}$ (\#2.7)
  magnetic space group.}

The magnetic unit cell doubles along $a$ and $c$ compared to the
nuclear one, as is shown in
Fig.~\ref{Fig_FePSe3_MagDiffraction_pressure}(c). Only $\rm Fe^{2+}$
atoms with opposing spin moments in black and white arrows are shown.
The nuclear unit cell is displayed with solid black lines as a
reference. We demonstrate the vdW layer in the bottom layer while
only showing the magnetic atoms within one cell for the remaining
layers. The moment direction is normal to the $ab$ plane, consistent
with the ambient-pressure data. The FM zigzag chains are along the $b$
direction, and are coupled with one another antiferromagnetically
within the plane.

Data at a constant counting time were also collected while warming
from the base temperature \replyv{(see Supplementary
  subsection~II.C)}. We were thus able to extract the temperature
dependence of the magnetic peak intensity at each pressure.  At
\unit[1.2]{GPa}, the intensity evolution was fitted to a
honeycomb-lattice 2D Ising
model~\cite{2007_Baxter_Models,2007_Rule_neutron_MagneticStructure_FePS3},
with $S$ set to 2. This fit yields an estimated Ne\'el temperature
$T_\mathrm{N}$ of approximately \unit[128]{K} (\replyv{Supplementary
  Fig~\ref{SuppFig_TN_Refinement1p2GPa}}\replyvs{details in the
  SI~\cite{Ref_SM}}), slightly higher than the previously reported
value of $108\sim119$~K at ambient
pressure~\cite{1981_Wiedenmann_neutron_MnPSe3_FePSe3,2023_FePSe3_PhononSymm_Spin-Phonon-Coupling}.
Data at \unit[4.1]{GPa} were too sparse for a meaningful fit but still
consistent with the trend of increased magnetic transition temperature
with elevating pressure.

The magnetic moment resulting from refinements was aligned parallel to
the \textit{c}-axis with a magnitude of
\unit[3.8(2)]{$\mu_{\mathrm{B}}$}.  The magnitude is lower than that
reported in Ref.~\cite{1981_Wiedenmann_neutron_MnPSe3_FePSe3}
($m_0=$\unit[4.9]{$\mu_{\mathrm{B}}$}).  We recognized that some of the
difference may be due to peak asymmetry, as the refinements in
Fig.~\ref{Fig_FePSe3_MagDiffraction_pressure} do not fully capture the
integrated intensities of the magnetic Bragg peaks.  It is possible
that the magnetic structure developed some disorder under pressure and
that some Bragg intensity was lost into diffuse scattering which was
subsequently removed using the background subtraction method.
\replyv{Supplementary} Fig.~\ref{SI_Fig:Neutron_Raw_BG_subtraction}
\replyvs{in the SI~\cite{Ref_SM}} shows some evidence for diffuse
scattering in the LP phase.

At \unit[8.3]{GPa}, at which point the crystal structure has already
changed to the HP phase with $R3$ symmetry, there was no signature of
any magnetic ordering at low temperatures, down to \unit[1.6]{K}.
Calculating the magnitude of scattering expected from typical Fe
moments and comparing with the size of signal observed in our previous
FePS$_3$ work~\cite{2021_Coak_MagPhase_FePS3}, we can be confident
that any signal would be visible above the level of noise in the data.
We cannot rule out the presence of magnetic diffuse scattering as this
might appear as background.  However, no clear intensity signal with
an obvious \textit{Q}-dependence was observed at any temperature
\replyv{(see Supplementary Fig.~\ref{SI_Fig:HP_temp_comparison})},
\replyvs{in the HP phase meaning that, as shown in the
  SI~\cite{Ref_SM},} \replyv{meaning that} any diffuse scattering
at \unit[8.3]{GPa} must be very weak and featureless.  This
implies that any surviving magnetic moments must be essentially
uncorrelated, leading to a paramagnetic phase.  We note that we cannot
distinguish a paramagnetic from a non-magnetic phase where Fe local
moments are lost.  However, we deem it unlikely for magnetism to
suddenly turn disordered down to \unit[1.6]{K} in the HP phase.

\subsection{Transport Measurements and Electronic Properties}

\begin{figure}[htbp]
    \centering
    \includegraphics[width=0.49\textwidth,angle=0]{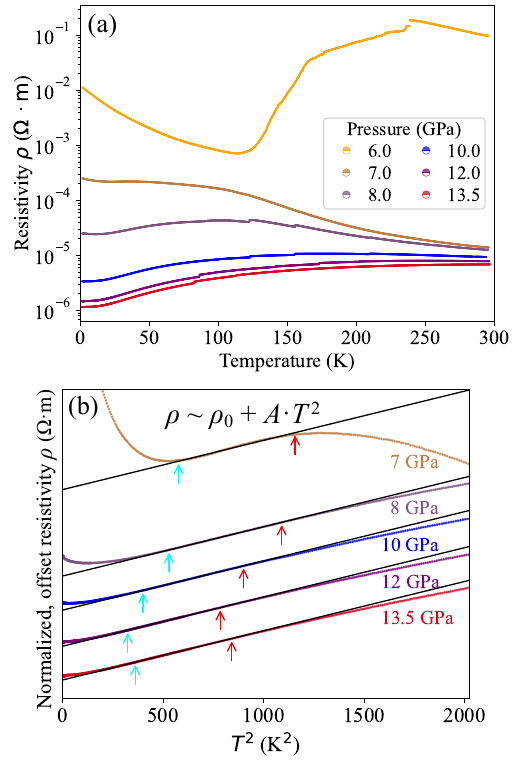}
    \caption{ \textbf{Single-Crystal Resistivity Measurements under
        Pressure} (a) The resistivity $\rho$ of a FePSe$_3$ single
      crystal, loaded in a Cubic Anvil Cell with glycerin as the
      pressure-transmitting medium.  (b) Normalized and offset
      resistivity as a function of $T^2$, a Fermi-liquid temperature
      power law, below $\sim$\unit[40]{K} for FePSe$_3$.  The fitted
      curves are plotted in solid black lines with the best-fitting
      $T^2$ region marked with arrows for each pressure.}
    \label{Fig_FePSe3_resistivity_pressure}
\end{figure}

Fig.~\ref{Fig_FePSe3_resistivity_pressure}~(a) presents the
temperature dependence of resistivity for a single-crystal FePSe$_3$
from 6.0 up to \unit[13.5]{GPa}.  The magnitude of the resistivity decreased
by four orders of magnitude at room temperature from 6.0 to \unit[7.0]{GPa},
indicating an insulator-to-metal transition (IMT).  Another
single-crystal sample measured from 2.0 up to \unit[14.0]{GPa}\replyvs{, shown
  in the SI~\cite{Ref_SM},} exhibited a consistent IMT between 6.0 and
\unit[8.0]{GPa} \replyv{(see Supplementary Fig.~\ref{SI:Fig_Resistance})}.

\subsubsection{Before the transition} 
Up to \unit[6.0]{GPa}, FePSe$_3$ exhibits an insulating character.  In
the high-temperature region ($T>$\unit[250]{K}), where the conduction
is dominated by thermally activated behavior, the data can be
well-described by an Arrhenius-type exponential expression, $\rho (T)
\sim e^{E_a / (k_{\scalebox{0.4}{B}} T)}$.  Plotting $ln(\rho)$
against $1/T$ allows us to perform a
\replyv{linear}\replyvs{polynomial} fit (detailed in the
\replyv{Supplementary subsection III.B.1} \replyvs{SI~\cite{Ref_SM}}),
which yields an extracted activation energy $E_a$ of 0.24, 0.21 and
\unit[0.11(3)]{eV} for 2.0, 4.0 and \unit[6.0]{GPa}.

At \unit[6.0]{GPa}, the resistivity gradient shows abnormal changes with a
change of sign in the intermediate temperature range which could
potentially result from underlying magnetic transitions or
Fermi-surface reconstruction in proximity to the LP to HP transition.
Further analysis and repeated thermal cycling would be needed to
uncover more details.  We also observed hysteresis in the \unit[6.0]{GPa} data
while cooling down and warming up, suggesting the IMT to be strongly
first-order.  Such hysteresis was not observed in data collected at
other pressures.

\subsubsection{After the transition}
\replyvs{Above \unit[6.0]{GPa} up to the highest pressure point, }We did not
observe superconductivity in two independent measurements using
different samples from the same batch for pressures up to
\unit[13.5$\sim$14.0]{GPa} and temperatures down to
$\sim$\unit[1.6]{K}. This was in contrast to a previous report of
superconductivity with a $T_c$ of \unit[2.5]{K} above
\unit[8.0]{GPa}~\cite{2018_Wang_pressure_SC_FePX3}. Our
crystallographic data showed our sample to be of good crystalline
quality and the structural model to be correctly interpreted.  Our
samples have equivalent resistivity magnitudes (and hence inferred
cleanliness) to those in Ref.~\cite{2018_Wang_pressure_SC_FePX3}. We
make further comments in the \replyv{Supplementary III.C.2.}
\replyvs{SI~\cite{Ref_SM}}.  In the main text, however, we focus on
the \replyvs{newly uncovered} exotic transport behaviors enriching the
\textit{T-P} phase diagram.

In the 7.0, 8.0, and \unit[10.0]{GPa} data, we find that the
resistivity increases initially upon cooling from \unit[300]{K},
similar to an activated behavior of an insulator or semiconductor.
The resistivity shows a broad maximum at approximately 40, 100,
and \unit[190]{K} for the 7.0, 8.0, and \unit[10.0]{GPa} data
respectively. Upon further cooling, the resistivity decreases showing
metallic behaviors. At higher pressure points (12.0 and
\unit[13.5]{GPa}) such a trend was not discernible within the measured
temperature range. The resistivity decreases upon cooling from
\unit[300]{K}.

Further analysis of the conducting phase finds that the resistivity
can be uniquely described by $\rho(T)=\rho_0 + A \cdot\ T^2$ at
temperatures below \unit[$\sim$40]{K}, as shown in
Fig.~\ref{Fig_FePSe3_resistivity_pressure}~(b).  We note that the
$A$-coefficient of the $T^2$ term decreases significantly with
increasing pressure\replyv{, as shown in Supplementary
  Fig.~\ref{SI:Resistitivy_HP_T2_LinerFit}}. \replyvs{The $T^2$ fits
  are detailed in the SI~\cite{Ref_SM}.}

This behavior, however, terminates with an upturn in the resistivity
at even lower temperatures. The low-temperature resistivity upturn is
observed around \unit[20]{K}, \unit[10]{K} and \unit[6]{K} in the 7.0,
8.0 and \unit[10.0]{GPa} data, respectively. The upturn is quickly
suppressed with elevated pressure, becoming completely absent above
\unit[10.0]{GPa}.

\begin{figure*}[htbp]
    \centering
    \includegraphics[width=0.99\textwidth,angle=0]{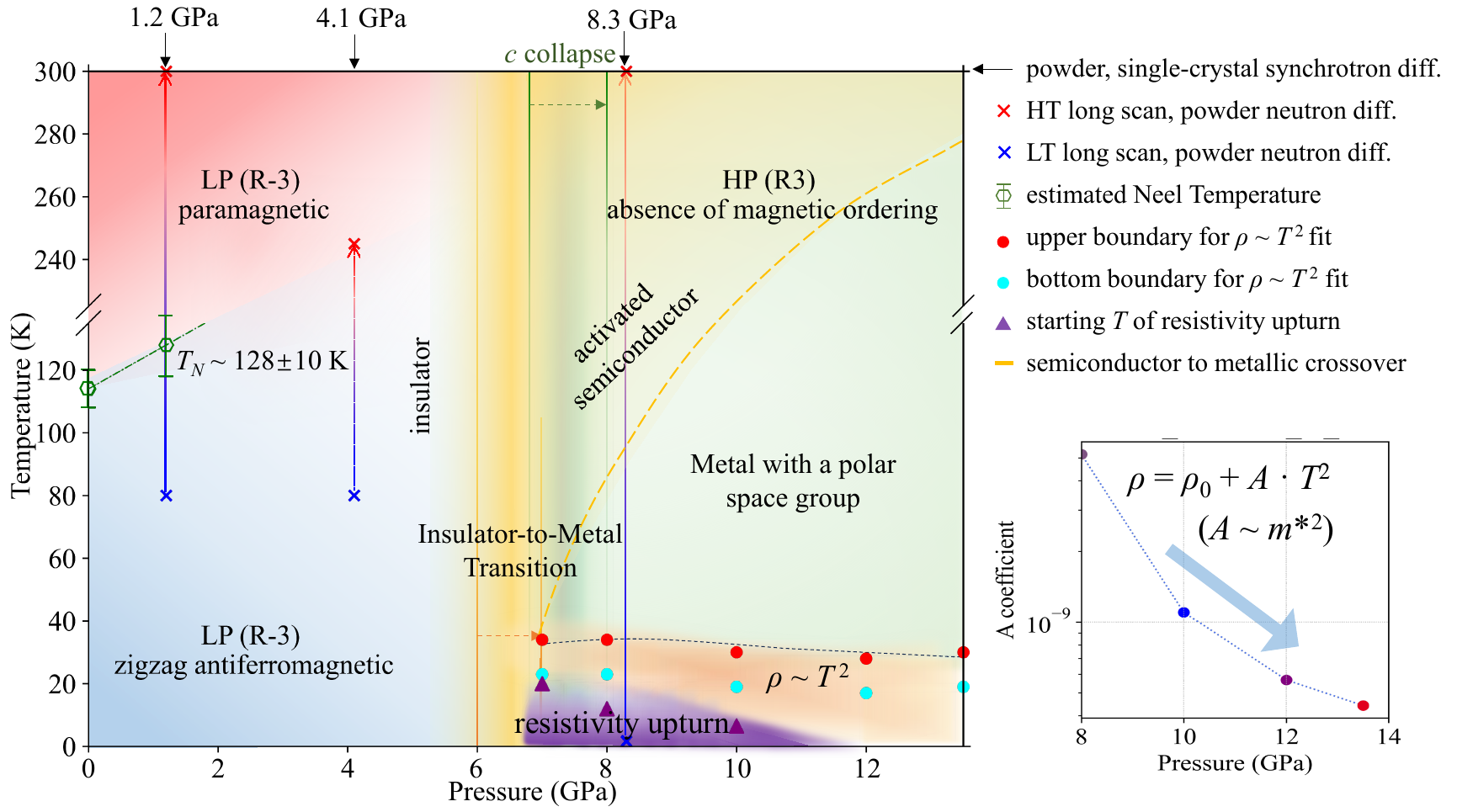}
    \caption{ \textbf{{\it T-P} phase diagram of FePSe$_3$ constructed
        from multiple probing techniques.}  The green open circles
      correspond to the antiferromagnetic transition temperature.  The
      one at ambient pressure is from literature (magnetic
      susceptibility measurement), and the one at \unit[1.2]{GPa} is
      derived from the intensity versus temperature change in our
      neutron diffraction data (fitted curve in the supplementary).
      The shaded light orange region refers to where the resistivity
      follows the $\rho \sim T^2$ behavior with the upper boundary in
      red, and the lower boundary in cyan. The region under the purple
      triangles is where the resistivity upturn was observed.  The
      inset at the bottom right shows the fitted coefficient $A$ of
      the $T^2$ term as a function of pressure for the shaded orange
      region.}
    \label{Fig_FePSe3_PhaseDiagram}
\end{figure*}
\section{Discussion}

Drawing together our results from multiple experimental probes, we are
able to complete the comprehensive temperature-pressure phase diagram
for FePSe$_3$ in Fig.~\ref{Fig_FePSe3_PhaseDiagram}. The color blocks
represent different phases, with gradient-shaded boundaries to guide
the eye.

Our synchrotron studies establish that FePSe$_3$ at \unit[300]{K}
undergoes a $R\overline{3}$ (LP) to $R3$ (HP) structural transition at
\unit[7$\sim$8]{GPa} of hydrostatic
pressure. \replyvs{During}\replyv{Across} this LP-HP transition, we
show a significant interplanar collapse along the {\it c}-axis,
\replyv{together with the absence of magnetic ordering from our
  neutron diffraction data}. The HP phase \replyvs{($R3$ space group)}
loses the unit cell's inversion center, leading to out-of plane
displacements of the Fe atoms\replyvs{ in the metallic phase}. The
unusual transition from a higher to lower symmetry space group under
pressure is unique in FePSe$_3$, making it different from other
compounds in this family.

Our resistivity measurements reveal exotic transport properties
changes in the phase diagram. The LP phase remains insulating up to
\unit[6]{GPa}, with the activated band gap shrinking in application of
pressure, compared to the ambient-pressure literature value, which
agrees qualitatively with the previous
report~\cite{2018_Wang_pressure_SC_FePX3}. We observe an
insulator-to-metal transition (IMT) between \unit[6 and 8]{GPa}, as
marked with the yellow dotted arrow in
Fig.~\ref{Fig_FePSe3_PhaseDiagram}.  The IMT transition pressure
corresponds to \replyv{that of} the LP-HP structural
transition\replyvs{ pressure}. We note, however, the high-temperature
region of the HP phase still shows activated semiconducting behavior.

At lower temperatures of the HP phase, FePSe$_3$ presents an
unexpected $T^2$ resistivity temperature dependence. The fitted
boundary of this region is marked with dots in
Fig.~\ref{Fig_FePSe3_PhaseDiagram}. It should be underlined that the
$A$-coefficient of the $T^2$ term decreases with increasing pressure,
indicating a substantial decrease in the effective mass of the charge
carriers.

An upturn in the resistivity, deviating from this state, emerges at
even lower temperatures, marked with purple. Such deviation is more
discernible in proximity to the AFM transition and is fully suppressed
at \unit[11$\sim$12]{GPa}. No bulk superconductivity was observed in
the HP phase up to the highest pressure (\unit[$\sim$14]{GPa}) in two
independent measurements of FePSe$_3$ samples from the same batch.

\replyv{Below we further discuss the key findings from this \replyvs{newly
  established} phase diagram, where the lattice, charge and spin orders
  evolution conindices and lead to novel phenomena.}

\replyv{First of all,} we identify that FePSe$_3$ undergoes an IMT
concomitant with the LP-HP structural transition. \replyvs{One aspect
  is that} \replyv{That is,} the associated interplanar {\it c}-collapse
coincides with the IMT.  Similar features were
previously reported in MnPS$_3$, FePS$_3$ and
NiPS$_3$~\cite{2018_Haines_highPressure_XRD_FePS3,2020_Coak_TuneDimensionality_TMPS3,2021_Ma_NiPS3},
but should not be considered as universal across this family, as
recent cases of VPS$_3$~\cite{2019_Coak_VPS3} and
NiPSe$_3$~\cite{2023_NiPSe3_Pressure_SC} exhibit Mott transitions
without structural changes.

Our key result is that, due to a periodic displacement of the Fe sites
as the system enters the HP phase, FePSe$_3$ becomes \replyvs{a polar
  metal}\replyv{metallic with local polar orders} at readily
accessible pressures (less than \unit[10]{GPa}). As this state can be
easily and cleanly tuned, the community now has a\replyv{n additional} 
\replyvs{new} handle on the
little-understood physics of these rare systems.  That is, the
material can be studied in both polar and non-polar configurations, at
all temperatures, and direct comparisons can thus be made.  In effect,
this allows a control to any tests, and lets us understand the initial
state that the \replyvs{new} physics emerges from.  \replyvs{We discuss more
  details in the following two aspects.}\\[6pt]

\begin{figure}
	\centering
	\includegraphics[width=0.49\textwidth,angle=0]{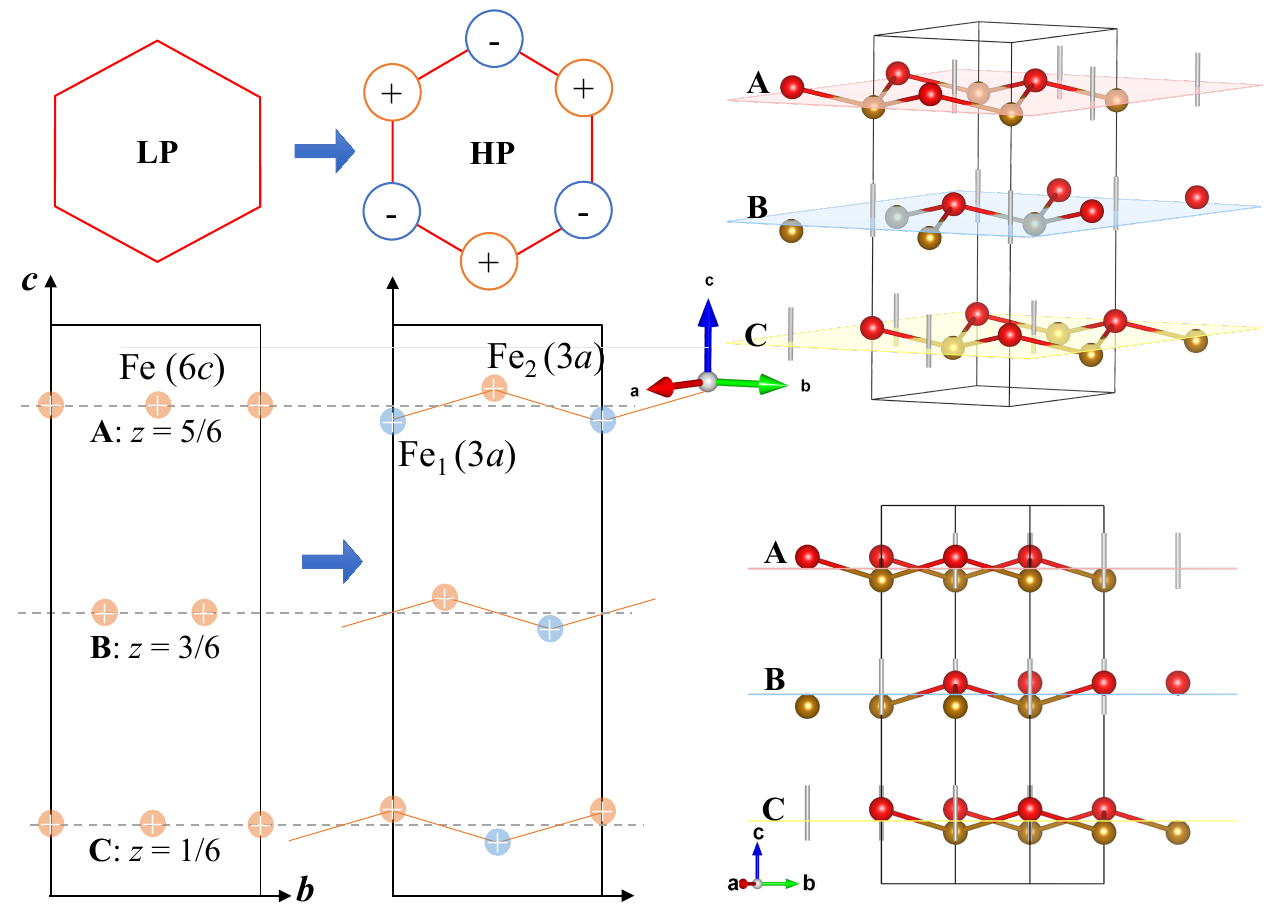}
	\caption{ \textbf{Cartoon of the Fe-hexagon buckling in the HP
            phase.}  The amplitude of the relative movement of Fe$_1$
          \replyv{(brown)} and Fe$_2$ \replyv{(red)} in reference to
          the $z$-centre of each plane has been increased for better
          eye guidance. Only Fe ions are shown here.}
\label{Fig_FePSe3_buckling}
\end{figure}

\replyv{Let us first look at the aspect of the crystalline symmetry.}
\replyvs{We identify the broken inversion symmetry and a polar point
  group symmetry in the HP structure from the synchrotron diffraction
  data.} In the initial LP phase, there is only one single
symmetry-constrained Fe site in the unit cell, and so all Fe atoms sit
\replyvs{exactly} on the $ab$ planes.  In the HP phase, however,
\replyv{we observed} the Fe-hexagon buckling\replyvs{has been observed
  in our results}, where two separate Fe sites are present in the
structure, as a result of an unusual lowering of symmetry under
compression.  As Fig.~\ref{Fig_FePSe3_buckling} demonstrates, the two
sets of iron \replyv{sites} move out-of-the-plane in opposing
directions\replyv{, denoted in different colors,} while the
\textit{z}-centre of the \textit{ab} plane remains as that in the LP
phase. \replyvs{We demonstrate the relative movement of one set of Fe
  moving up and one moving down in different colors.}

Our detailed analysis of the atomic displacements suggests local net
dipole moments to form in the HP phase along the $c$-axis due to the
Fe buckling.  Additionally, we observed relative displacement of the
Se and P atoms but the amplitude is rather subtle.  We believe further
investigations are needed to identify whether the system stabilizes in
an anti-ferroelectric state or could potentially allow for a net
dipole moment forming as a result of ligand atom displacement.
To date, polar metals can be divided into those that are essentially
ferroelectric semiconductors and have significant carrier densities
doped into the system to give a weakly metallic state or metals where
no bulk polarization (effectively capacitance) is measurable as the
free carriers move and reconfigure to screen any microscopic dipoles.
The latter case results in essentially a periodic modulation of
electron clouds or potentials conceptually similar to the
perturbations to electronic transport seen in a charge density wave
(CDW), but built on totally different foundations.  The electric
dipoles in polar metals arise as driven by lattice change, whereas a
CDW arises from electronic Fermi surface nesting.  FePSe$_3$ is unique
in being tuned from insulator to semiconductor to metal with pressure,
whilst \replyvs{at the same time} the polar point group develops with
the onset of the semiconducting state.  This presents an opportunity
to understand how the screening develops as a function of carrier
concentration, free from disorder effects in cases of chemical doping.
Importantly, partially screened dipoles are an intriguing candidate
for unconventional superconductivity, \replyv{as previously seen}
in the doped polar metallic state of SrTiO$_3$~\cite{bhowal_polar_2023}.
Indeed, the possibility of strong
electron-electron correlations mediated by unscreened or
under-screened dipoles (static or fluctuating) presents an exciting
opportunity for \replyvs{new}\replyv{unknown} superconducting materials.

Polar metals, combining two phenomena usually considered mutually
exclusive, are \replyv{promising multifunctional
  materials~\cite{2023_PRM_MJames_PolarMetals} and are} increasingly
viewed as topologically significant
systems~\cite{bhowal_polar_2023}. \replyvs{The physics of these
  materials is yet in their infancy and holds great promise.}
Though\replyvs{ polar metals are promising multifunctional
  materials~\cite{2023_PRM_MJames_PolarMetals},} predicted back in the
60s by Anderson~\cite{Anderson_PolarMetals_1965}, they have only
recently started to be demonstrated in very few materials,
\replyv{among which} 2D examples are \replyv{even} rarer\replyvs{
  still}. So far there are only a few known examples:
CePt$_3$Si~\cite{2004_PRogl_PRL_Noncent}, UIr~\cite{2004_UIr_JPSJ},
LiOsO$_3$~\cite{2013_FE_transition_in_metal},
WTe$_2$~\cite{2019_RoomT_FE_semimetal}, doped
SrTiO$_3$~\cite{2020_Corentin_SC_polar_modes_FE} and pressurized
Hg$_3$Te$_2$X$_2$~\cite{2021_Pressure_FE_tran_Polar_Metal}. We note
that NiPS$_3$ has been recently suggested to be another physical
system under high pressure (above
\unit[$\sim23$]{GPa})~\cite{2022_NiPS3_pressure_symmetry_PolarMetal}.
\replyvs{But their}\replyv{However, the} limited access to
\replyvs{room-temperature} measurements \replyv{under such high pressures}
makes it difficult to go beyond speculating the possibility of such phases.
FePSe$_3$\replyv{, however, provides the field a promising
  and tunable example at moderate pressure values.}

\replyvs{As the polar metal state can be induced and then tuned with
  pressure, FePSe$_3$ gifts the community a new handle on the
  little-understood physics of these exotic systems.  That is, the
  material can be studied in both polar and non-polar configurations
  at all temperatures, including intermediate phases showing
  nontrivial transport properties, thus allowing for direct
  comparisons.}

 Our unambiguous structural model of this clean and simple system lays
 a solid foundation for further experiments and density functional
 theory calculations, and we identify multiple questions raised by the
 anomalous transport data in this \replyvs{new }phase for ongoing focused study
 in the field.\\[2pt]

\replyv{The second aspect lies in the electronic transport properties.}
The HP polar phase involves rich phases in the \textit{T-P} phase
diagram, far from being a simple metal.  In the high-temperature
region\replyvs{ of the polar HP phase}, FePSe$_3$ shows activated
semiconductor behavior instead of turning metallic immediately, with
the resistivity value in the order of 10$^{-4}~\Omega \cdot m$.  The
crossover \replyvs{temperature} between the high-temperature
non-metallic and lower temperature metallic region at varying pressure
values forms a fan-shaped region\replyvs{ around the IMT}. It suggests
another interesting intermediate state for further exploration.

Below \unit[$\sim40$]{K}, we observe that the resistivity\replyvs{ of
  FePSe$_3$} demonstrates a $T^2$ region, which distinguishes it from
other compounds of the family. We note that the fitted
$A$-coefficients of the $T^2$ term are not constant in the HP phase
but rather decrease with increasing pressure
(Fig. \ref{Fig_FePSe3_PhaseDiagram}).  If we were to take the
quasi-particle description, where the $A$-coefficient reflects the
effective mass of electrons, such a trend suggests that the electrons
are turning less correlated and electron screening effects getting
enhanced with increasing pressure.  However, this `good metal'
Fermi-liquid description cannot reasonably be applied to a compound
with such high resistivity and residual resistivity.  Carriers are
being heavily scattered, by some alternative (evolving) mechanism that
is yet to be determined, but is most likely linked to the polar state,
the low dimensionality, or indeed both.

At even lower temperatures, the resistivity deviates from the $T^2$
behavior and shows an upturn.  Similar deviations have been previously
reported in other $TM$P$X_3$ materials and are still awaiting full
explanations~\cite{2018_Haines_highPressure_XRD_FePS3,2020_Coak_TuneDimensionality_TMPS3}.
In the case of FePSe$_3$, it should be noted that the upturn is
quickly suppressed with increasing pressure, showing the same trend as
the $A$-coefficient shrinking with pressure.  It remains an open
question whether \replyvs{the phase where }this \replyv{nontrivial}
upturn \replyvs{dominates the conductivity }(purple region in
Fig.~\ref{Fig_FePSe3_PhaseDiagram}) is related to the
antiferromagnetic critical transition pressure point, or the localized
polar modes taking effect when the electrons have the smallest
mobility values and are nearly free from thermal effects.

More generally, the strongly correlated electrons are likely to
introduce quantum critical fluctuations close to any quantum critical
point or endpoint associated with magnetic
transitions~\cite{2000VThomas_PRL,2004_UIr_JPSJ}, leading to emergent
phenomena including unconventional superconductivity or novel metallic
states. The LP-HP structural transition is strongly first-order, so we do
not have clear smoking gun for such physics, but hints of more
subtle changes in order lying behind the phase diagram.  Additionally,
the broken inversion symmetry may yield a Rashba-type spin-orbit
coupling~\cite{2004_PRogl_PRL_Noncent} and thus exclude spin-triplet
pairing in the HP phase of FePSe$_3$.  Further investigations are
needed to deepen the understanding of this non-centrosymmetric
compound's novel phases and underlying mechanism.

The \replyv{exotic} transport properties contrast FePSe$_3$ from the
rest of the $TM\rm{P}X_3$ family and merit further investigations into
the many questions \replyv{posed above} \replyvs{they pose}. Future
work based on these discoveries should provide many insights into the
fundamentals of van-der-Waals strongly correlated magnetic compounds
and into polar metal states.\\[2pt]

\replyv{Now, let us look further into the correct interpretation of
  the crystalline transition in van der Waal compounds.}  We note that
our results showing the LP-HP transition model with essential
interplanar lattice collapse contrasts with the intraplanar lattice
collapse interpretation previously proposed by Wang
{\it et al.}~\cite{2018_Wang_pressure_SC_FePX3}. We would like to
underline the fact that crystallography in layered materials has been
challenging due to issues such as preferred orientation of lamellar
powder grains and stacking faults along the interplanar axis. These
issues turn more profound in the application of pressure, where the
diffraction data collection angles are constrained. Similar cases
have been previously reported\replyvs{ in this compound family}, such as in
NiPS$_3$~\cite{2024_MPX3_Pressure_review_Mandrus}. We thus found it
essential to optimize the methodology and to draw on as wide a
selection of experimental techniques as possible, so as to reach
solid, unambiguous structural conclusions. This is reflected in the
following aspects:

\replyvs{\textbf{Using as-grown powder samples}}
\replyv{(1)} We used powder samples from growth to minimize preferred
orientations,
instead of the usual way to grind single crystals to
powder. Ref.~\cite{2018_Wang_pressure_SC_FePX3} used pre-compressed
FePSe$_3$ pellets and silicone oil as a PTM for data collection.

\replyvs{\textbf{Optimized collection geometry - }} \replyv{(2) We
  used a large wavelength ($\lambda$=0.4246~{\AA} for powder;
  $\lambda$=0.4859~{\AA} for single-crystal) to ensure a sufficiently
  wide $Q$ range to resolve the (003) peak.  In contrast,
  Ref.~\cite{2018_Wang_pressure_SC_FePX3} was not able to access the (003) peak
  due to the much smaller wavelength (\unit[$\lambda$=0.3100]{\AA}, powder
  only). Their tracked (006) peak in fact overlaps with the (015) and
  (110) reflections in a hexagonal cell, mixing up both intra- and
  inter-planar lattice parameters.} \replyvs{The key difference
  between our work and that of Wang
  $et.\ al.$~\cite{2018_Wang_pressure_SC_FePX3} is that we have a
  sufficiently wider $q$ range to resolve the (003) peak.}  Tracking
the $2\theta$ angle where the \replyv{(003)} peak is observed in
powder diffraction data gives us direct and model-independent evidence
of the interplanar $d$-spacing evolution.  The most intense (113)
peak, however, reflects changes in both inter- and intraplanar
distances and thus cannot give solid conclusions.

\replyvs{\textbf{Effect of pressure-transmitting medium - }}
\replyv{(3)} We additionally examined the \replyv{effect of the}
presence of a PTM.  We utilized helium as it has been documented to be
the best available PTM, even in its solid state above
\unit[$\sim$12.1]{GPa}~\cite{2009_Klotz_HydrostaticLimits_PressureMedia}.
We found that, in the absence of a PTM, the onset of the LP-HP
structural transition was observed at a higher pressure value
(\unit[$\sim$9.6]{GPa}), and was followed by a co-existence region of
both LP and HP until \unit[$\sim$16.3]{GPa} \replyv{(see Supplementary
  Fig.~\ref{SI:Fig_PowderDiff_noH})}.  A pure HP phase was not evident
before this pressure value,\replyvs{ as detailed in the
  SI~\cite{Ref_SM}} \replyv{but the LP-HP transition itself is robust}.

\replyvs{\textbf{Combined powder \& single-crystal diffraction - }}
\replyv{(4) Lastly, we have to underline} \replyvs{we found} that powder
measurements themselves are not enough to conclude the structural
transition model.  It was the single-crystal synchrotron diffraction
that gave a solid solution. In our case, \replyvs{the
  model-independent proof is given through} the ($h\overline{h}l$)
plane slices (Fig.~\ref{Fig_FePSe3_Xray_LatticeChange}(b)) give us
\replyv{the model-independent proof of} an interplanar collapse
\replyvs{occurred} during the LP-HP transition. After the transition,
we discern a substantial breaking of the Friedel pairs' structure
factors, which can only be attributed to the broken inversion
symmetry. \replyvs{ It should be noted that} Such a symmetry change is
not discernible by powder diffraction.

We thus confidently affirm our LP-HP structural transition model.
Notably, this is an intuitive finding as the planes joined merely by
weak van der Waals forces are much easier to compress than it is to
compress the hexagonal network of covalent bonds within the layers. We
have additionally developed a methodology to carry out the data
collection in the most robust way to address the challenge in solving
the crystallography in low-dimensional materials.\\[2pt]

\replyv{In addition to the lattice and charge perspectives, let us now
  look at the magnetic order transition.}
Our powder neutron diffraction studies \replyvs{under pressure} affirm
that the LP phase experiences a paramagnetic to antiferromagnetic
transition \replyv{with decreasing temperature}. The magnetic
configuration of LP remains the same as that reported in the
ambient-pressure~\cite{1981_Wiedenmann_neutron_MnPSe3_FePSe3}, with
$k=(\nicefrac{1}{2},0,\nicefrac{1}{2})$ and the moments parallel to
the $c$-axis \replyv{in the magnetic space group $P_s\overline{1}
  (\#2.7)$}. Zigzag ferromagnetic chains form within the vdW planes
and couple with one another antiferromagnetically.  The
\replyv{nuclear} space group symmetry remains unchanged with varying
temperature.

\replyv{With increased pressure in the LP phase, the
  magnetic order remains.} Yet, the magnitude of the magnetic moment is
suppressed\replyvs{ with increased pressure}, potentially due to
emergent diffuse scattering. The magnetic transition temperature $T_N$
increases slightly\replyvs{ with elevated
  pressure}. \replyvs{In the HP phase, we did not observe any presence
  of magnetic ordering from 300~K to the base temperature 1.6~K.}

\replyvs{We find that the magnetic configuration of LP is consistent with the
ambient-pressure model~\cite{1981_Wiedenmann_neutron_MnPSe3_FePSe3},
while the size of the magnetic moments shrinks with increased
pressure.  Additionally, the value of $T_\mathrm{N}$ tends to increase
with elevated pressure.}  We note that the interplanar distance shrinks
more rapidly than the intraplanar atomic separation in the LP phase,
which might affect the coordination around the magnetic fragments
subtly.  Further explorations, for instance, magnetic susceptibility
measurements, neutron inelastic scattering under pressure or magnetic
exchange interaction simulations, would be needed to gain full
understanding into the magnetic phase transitions.

After the structural transition in the HP phase, however, no magnetic
peaks or short-range-ordering features \replyv{or intensity change in
  the nuclear peaks} are visible\replyvs{ in our neutron powder diffraction
data.}  We thus conclude that FePSe$_3$ does not order magnetically in
the HP phase, in interesting contrast to the sulfur compound
FePS$_3$~\cite{2021_Coak_MagPhase_FePS3}.  The system may experience a
reconfiguration of electronic states as FePSe$_3$ becomes more
three-dimensional and carriers become itinerant after the LP-HP
transition.  Though we cannot distinguish whether the HP phase is
paramagnetic or non-magnetic from our powder neutron diffraction data,
we however deem it unlikely for magnetism to stay disordered down to
\unit[1.6]{K}.  We thus strongly favor the latter interpretation that
FePSe$_3$ turns from high-spin ($S$ = 2) to non-magnetic ($S$ = 0)
state upon metallization, which also is in line with the X-ray
emission spectroscopy observation by Wang
\textit{et al.}~\cite{2018_Wang_pressure_SC_FePX3}.


To conclude, we tune the van-der-Waals compound FePSe$_3$, which is a
Mott-insulator with antiferromagnetic ordering at ambient pressure, to
a high-pressure polar phase showing semiconducting features at higher
temperature, and metallicity at lower temperatures via the application
of pressure.

We identify a crystalline transition from the centrosymmetric
$R\overline{3}$ to the non-centrosymmetric $R3$ space group at
approximately $7$~GPa through rigorous synchrotron diffraction
studies.  This transition is accompanied by a substantial interplanar
lattice collapse with no evidence of interplanar sliding.  This is \replyvs{the
first}\replyv{an} unambiguous structure transition model in FePSe$_3$ and presents
a unique transition towards a higher-order state under pressure.  Most
importantly, the local distortion of Se atoms and the buckling of the
Fe hexagons at the loss of inversion symmetry when the material
enters the HP structural phase causes the formation of a polar axis
and local dipoles.  We also observe the suppression of magnetic
ordering and an insulator-to-metal transition at the same pressure,
the LP-HP structural transition.

Based on the symmetry analysis and change of conductivity, we
\replyvs{report}\replyv{suggest} the emergence of a polar metal phase
in FePSe$_3$ at accessible pressures.  \replyvs{Polarization and
  metallicity have been assumed to be mutually exclusive, until the
  recent few experimental realizations.}  Our findings thus provide a
\replyvs{new} platform to explore polar metallicity, in a chemically simple and
clean 2D system with strongly correlated electrons.  The 2D nature of
FePSe$_3$ also offers the possibility to control the polarization
along the $c$-axis while maintaining the anisotropic metallicity in
the $ab$ plane.  Further investigations, beyond the scope of this
work, will be needed to explore the polarization and charge
distributions in this \replyv{potential} polar metallic phase - an
exciting opportunity \replyvs{for discovery }in this \replyvs{new }field.

Additionally, we outline intermediate phases manifesting exotic
metallic behaviors near the structural and magnetic transition
pressure. A $T^2$ resistivity exponent suggests at first glance a
simple Fermi-liquid metal (which in itself would be an interesting
contrast to the behavior seen in FePS$_3$
\cite{2020_Coak_TuneDimensionality_TMPS3}). However, both the
resistivity magnitude and the residual resistivity exhibited place
this weakly conducting state well outside the range where a Fermi
liquid picture could be applied. The scattering mechanisms must be
something different; more microscopic and electronics probes will be
needed to ascertain exactly what their fundamentals are.  Finally, the
suppression of the coefficient of the $T^2$ term at low temperatures
and the suppression of a resistivity upturn below the $T^2$ region,
appearing only close to an AFM critical transition pressure,
strongly suggests some underlying quantum ordering or remnant
disorder, necessitating further exploration.

Overall, our multi-probe studies of the crystalline, magnetic
structures and transport properties of FePSe$_3$ contribute
significantly to the intrinsic understanding of the $TM\rm{P}X_3$
compound family.  This work presents \replyvs{the first} \replyv{a}
fundamental understanding of the high-pressure structures and the
\replyvs{discovery of a new} emergent 2D
\replyvs{polar} metallic phase \replyv{with a polar space group}.
These findings offer a solid starting point for further physical
properties investigations and {\it ab-initio} studies, providing
insights into the fundamentals and tunability of van-der-Waals
strongly correlated magnetic compounds and presenting a uniquely
tunable low-dimensional model system for the \replyvs{newly-}emerging
field of polar metallicity.  Further study, involving the control of
individual layers and stacking patterns, will also be an area of
interest for advancing the engineering of low-dimensional magnetism in
van-der-Waals materials and applying them into atomically thin-layer
device architectures.

\section*{References}
\bibliographystyle{naturemag}
\bibliography{ref_aps}

\section{Methods}

\subsection{Sample Synthesis and Characterization}

FePSe$_3$ samples were synthesized via two synthesis routes. Both
routes had starting materials sealed in quartz tubes that had been
cleaned by an acid etching followed by rinsing with demineralized
water.  All starting materials were of 99.99\% purity.  Powder
samples, used for x-ray and neutron diffraction, were made using a
solid-state reaction. This route was used to minimize preferred
orientation, known to occur when grinding single crystals due to the
vdW nature of the compound.  An equivalent of  \unit[2]{g} of starting
materials were sealed in quartz tubes, with stoichiometric quantities
of Fe and Se.  The P content was 10\% larger than stoichiometric, with
the excess included to account for possible impurity phases.  The
tubes were sealed containing a pressure of \unit[5]{torr} of argon. The tubes
were then placed in a horizontal two-zone furnace with the starting
materials in zone 1. The zones were independently heated to follow the
temperature protocol listed in Table~\ref{tab_HeatProt}. The protocol
prevented vapor transport as zone 1 was always cooler than zone 2.

\begin{table}[htbp]
\centering
\caption{
\textbf{Temperature protocols during sample synthesis.}
Temperature protocols used to synthesize the powder and single crystal samples of FePSe$_3$.}
\vspace{6pt}
\begin{tabular}{cc@{\quad}c@{\quad}c}
\hline \hline 
    & Zone 1 ($^\circ$C) & Zone 2 ($^\circ$C)  & Duration (h) \\
\hline
    \multirow{3}{*}{Powders} 
    & 25 $\rightarrow$ 650 & 25 $\rightarrow$ 670 & 6 \\
    & 650 & 670 & 168 \\
    & 650 $\rightarrow$ 25 & 670 $\rightarrow$ 300 & free cooling \rule{0pt}{2.5ex}{\hfill} \\ 
\hline
    \multirow{5}{*}{Single Crystals} & 25 $\rightarrow$ 750 & 25 $\rightarrow$ 720 & 6 \\
    & 750 & 720 & 168 \\
    & 750 $\rightarrow$ 500 & 720 $\rightarrow$ 450 & 48 \\
    & 500 $\rightarrow$ 150 & 450 $\rightarrow$ 120 & 6 \\
    & 150 $\rightarrow$ 25 & 120 $\rightarrow$ 25 & free cooling \\
\hline \hline 
\end{tabular}
\label{tab_HeatProt}
\end{table}

The two-zone furnace was also used to synthesize single crystals of
FePSe$_3$. Equivalent of \unit[1]{g} of starting material, consisting of
stoichiometric quantities of Fe, P and Se, were placed in quartz
tubes.  A small amount of iodine (\unit[0.05]{g}) was included to facilitate
vapor transport.  The tubes were then sealed under vacuum and loaded
in the furnace.  The temperature protocol is also listed in
Table~\ref{tab_HeatProt}. Zone 2 was colder than zone 1 and the
crystals grew in zone 2 via a vapor transport mechanism. The crystals
were recovered at the end of the synthesis.

\subsection{Synchrotron Diffraction under Pressure}

Powder diffraction data were collected at room temperature on the
Extreme Conditions beamline (I15) at Diamond Light Source, UK.  An
incidence X-ray beam with focused spot size \unit[30]{$\mu m$} and
wavelength \unit[0.4246]{\AA} (equivalently to \unit[$E$=29.2]{keV})
was used to collect the diffraction patterns.  A MAR345 area detector
was used to record the diffraction pattern with an exposure time of
120 seconds for each scan at a distinctive pressure value.  The
instrument parameters were calibrated with LaB$_6$.  The data were
initially processed using the data analysis software {\tt
  DAWN-II}~\cite{2017_Filik_DAWN2}.  Further, Le Bail and Rietveld
refinements were performed using {\tt
  GSAS-II}~\cite{2013_Toby_GSAS-II}.

Single-crystal diffraction data were collected at room temperature on
the Small Molecule Single Crystal Diffraction beamline
(I19-2)~\cite{2012_I19_Nowell} at Diamond Light Source, UK.  Samples
were manually reduced to a thickness of about \unit[10]{$\mu m$} and
squares of side length \unit[80]{$\mu m$}.  The plate-like sample thus
had the $c \parallel c^*$ axis, pointing perpendicular to the crystal
planes.  The sample was laid flat on an anvil.  Thus, $c^*$ was
parallel to the anvil axis and close to the direction of the incident
synchrotron beam. The experimental geometry is comprehensively
described in the Supplementary
\replyv{Fig.~\ref{SI:Fig_Synchrotron_SingleCrystal_Scheme}}. An
incident beam of wavelength \unit[0.4859]{\AA} was used and
diffraction data were collected while the crystal rotated in a
standard configuration with a Dectric Pilatus \unit[300]{K}
detector. The data were then analyzed using {\tt CrysAlis Pro}~\cite{Agilent_CrysAlisPro}.
\replyv{A numerical analytical absorption correction was applied based
  on a multifaceted crystal model using a calculated linear absorption
  coefficient of \unit[$\mu$=7.720]{$~mm^{-1}$}, combined with an
  empirical correction via the {\tt SCALE3-ABSPACK} algorithm,
  yielding transmission factors of 0.843--0.857.}

Diamond anvil cells (DAC) with rhenium gaskets were used for both
powder and single-crystal samples.  The culet size of a DAC was
approximately \unit[400]{$\mu m$}.  A hole was drilled through a
rhenium gasket of a diameter less than half that of the culets.  The
gasket hole contained the FePSe$_3$ sample, and some small ruby
crystals used as a pressure gauge.  Two DACs were prepared for powder
and single crystal diffraction experiments, both being loaded with
helium gas serving as the pressure-transmitting medium (PTM).

Pressure was applied using a gas-loading membrane system.  The
generated pressure inside the sample space was determined by optical
measurement of the characteristic $R_1$ fluorescence peak of
ruby~\cite{1986_Mao_Ruby_Pressure_calibration,2020Shen}.  The peak
position was measured before and after data collection at each
pressure point and the average value was taken as the estimated
pressure during the measurement.  The uncertainties in the pressure
values were estimated to be approximately \unit[$\pm0.1$]{GPa}.

Diffraction peaks from the DAC were identified by fitting the known
diamond unit cell and were removed for subsequent treatment.
Powder-like diffraction rings from the rhenium gasket were also
identified and excluded from refinements.  Details of pressure
determination, sample data reduction and analysis are in the
\replyv{Supplementary Section I.~B-E}.

\subsection{Powder Neutron Diffraction under Pressure}

Powder neutron diffraction patterns were collected on the D20
instrument~\cite{2008_Hansen_D20} at the Institut Laue-Langevin,
France~\cite{2021_ILL_FePSe3MagExp_DOI}, in a comparable manner to our
previous study on FePS$_3$~\cite{2021_Coak_MagPhase_FePS3}. The
neutron wavelength was \unit[2.42]{\AA}, set via a graphite
monochromator. Data were collected using a Paris-Edinburgh type
pressure cell press~\cite{Klotz2005} with double-toroidal sintered
diamond anvils~\cite{Klotz2016}. The applied load on the cell was
dynamically controlled to be constant as a function of temperature.

Powder samples from the same batch used in the synchrotron powder
measurements were used for this experiment. The powder was packed into
the two halves of a Ti/Zr null matrix spherical
gasket~\citep{Marshall2002} and then placed between the anvils.  A
{4:1 deuterated methanol/ethanol} mixture served as a hydrostatic
pressure medium. No pressure gauge was employed, as the pressure
dependence of the sample's lattice parameters (at room temperature)
was taken from the synchrotron diffraction results. The pressure was
determined by refining the neutron diffraction patterns and extracting
lattice parameters, which were then matched to the synchrotron data.
While the synchrotron data were measured at ambient temperature, no
great change in the positions of the nuclear Bragg peaks were observed
as a function of temperature in the neutron data. Consequently, the
uncertainty in sample pressure was estimated to be
\unit[$\pm$0.3]{GPa}.  All pressure changes were made at temperatures
where the pressure medium was known to be in its liquid
phase~\cite{2009Klotz}.

Table~\ref{tab_MagNeutron} summarizes the scans for the different
pressures taken at high- (HT) and low-temperature (LT).  At HT, data
were collected with 60-minute count times at 300, 300, 245 and \unit[300]{K}
for pressure values of 0, 1.2, 4.1 and \unit[8.3]{GPa}, respectively.
Subsequently, the cell was rapidly cooled to 80~K under constant load
by being immersed in liquid nitrogen.  At the highest pressure, the
cell was subsequently cooled from 80 to \unit[5]{K} using a closed cycle
cryocooler and further cooled to \unit[1.6]{K} by immersing the cell into
liquid $^4$He and pumping on it.

\begin{table}[htbp]
\centering
\caption{
\textbf{Key scans of magnetic scattering data.}
Scans used in the main text for exploring the magnetic properties evolution in powder samples of FePSe$_3$.}
\vspace{12pt}
\begin{tabular}{c@{\quad}c@{\quad}cc@{\quad}c@{\quad}c}
\hline \hline 
\multirow{2}{*}{\makecell{Pressure\\(GPa)}} & \multicolumn{2}{c}{HT} &  &\multicolumn{2}{c}{LT} \\
\cline{2-3} \cline{5-6}
    & T (K) & Count Time & & T (K) &  Count Time \\
\hline
0   & 300 & 60~min & & \multicolumn{2}{c}{--} \\
1.2 & 300 & 60~min & & 80 & 8$\times$15~min\\
4.1 & 245 & 60~min & & 80 & 8$\times$15~min\\
8.3 & 300 & 60~min & & 1.6 & 8$\times$30~min\\
\hline \hline
\end{tabular}
\label{tab_MagNeutron}
\end{table}

The raw diffraction data had a broad, diffuse background due to the
pressure environment and liquid pressure medium.  An independent
background from an empty cell could not be accurately measured because
of uncertainties positioning the cell and equating the same volume of
pressure medium as well as changes in the background scattering due to
temperature and pressure.  Consequently, the background was estimated
by manually specifying background points away from any diffraction
peaks and fitting a smooth Chebychev polynomial through them to
subtract from the raw data.  The method allows for Bragg peaks to be
indexed and their intensities to be determined, but could not
discriminate against any diffuse scattering that may come from the
sample.  All intensities are normalized to a monitor placed in the
incident beam.  The resulting data were then refined using the
Rietveld method as implemented in software {\tt
  GSAS-II}~\cite{2013_Toby_GSAS-II} and {\tt
  FullProf}~\cite{2001_FullProf}.  More details on data processing can
be found in the \replyv{Supplementary subsection II.A and
  II.B}.

\subsection{Resistivity Measurement with Cubic Anvil Cell}
The resistivity of a FePSe$_3$ single crystal sample, from the same
batch used for synchrotron diffraction, was measured under various
pressures at the Cubic Anvil Cell (CAC) station of Synergetic Extreme
Condition User Facility (SECUF), China~\cite{2018_Cheng_CAC_HP_LT}.
The sample was prepared with dimensions \unit[0.23]{mm} in length
(\textit{l)}, \unit[0.22]{mm} in width (\textit{w}) and
\unit[0.01]{mm} in height (\textit{h}).  A standard four-probe method
was used for resistivity measurements under high pressure.  Silver
paste (DuPont 4929N) was used to glue gold wires, \unit[20]{$\mu m$}
in diameter, to the sample.  The resistance ($R$) of the sample was
measured, and the intrinsic resistivity $\rho$ was calculated using
the formula, $\rho\ =\ R* (w \cdot h)/l$.

The sample was placed inside a Teflon capsule filled with glycerol PTM
before being loaded into the CAC.
\replyvs{The three-axis compression geometry together with the adoption of a liquid PTM}
\replyv{A CAC is known to provide better hydrostatic conditions than a
  conventional opposed-anvil DAC, due to its three-axis compression
  geometry~\cite{2006_Sterer_cDAC}. The glycerol shows good
  hydrostaticity up to high
  pressures~\cite{2009_Klotz_HydrostaticLimits_PressureMedia}.
  Combining these two aspects thus} ensured
pressure homogeneity
\replyv{comparable to those using helium as a PTM
  in a conventional DAC}. The pressure values in the CAC were
estimated from a pressure-loading force calibration curve
pre-determined at room temperature. Temperature-dependent
measurements from 300 to \unit[1.6]{K} were performed using a liquid
$^4$He cryostat with a \unit[9]{T} superconducting magnet.  A helium
exchange gas was used to equilibrate the temperature of the CAC. A
thermometer was inserted inside the CAC to keep good thermal contact.
Details about assembling samples and calibrating the pressure of a CAC
can be found in Ref.~\cite{2014_CubicAnvil_HP}.

At the lowest pressure point (\unit[6.0]{GPa}), where the sample was expected
to be an insulator, two methods were used to measure \textit{R}. From
300 to \unit[215]{K}, \textit{R} was measured using a Lake Shore Model
372 in alternating current (AC) mode with a current of
\unit[10$^{-7}$]{A} at a frequency of \unit[13.7]{Hz}. From
\unit[215]{K} down to the base temperature \unit[1.6]{K}, data were
collected with a Keithley 2400 and 2182 in direct current (DC) mode
with an average current of \unit[10$^{-7}$]{A}. For higher pressure
points, where the sample was expected to be a metal, the data were
collected using a Keithley 2400 and 2182 in DC mode with an average
current of \unit[10$^{-5}$]{A} for 7.0 and \unit[8.0]{GPa}, and
\unit[10$^{-4}$]{A} for 10.0, 12.0 and \unit[13.5]{GPa} from 300 to
\unit[1.6]{K}.

\section{DATA AVAILABILITY}

\replyv{The powder and single-crystal synchrotron diffraction data are
  available in the DLS Data Gateway
  (\href{https://icat.diamond.ac.uk}{https://icat.diamond.ac.uk}),
  with the proposal number CY31687-1 and CY31687-2.  The powder
  neutron diffraction data are available in the ILL Data Portal
  (\href{https://data.ill.fr/}{https://data.ill.fr/}), with the
  reference
  \href{https://doi.ill.fr/10.5291/ILL-DATA.5-31-2847}{DOI:10.5291/ILL-DATA.5-31-2847}~\cite{2021_ILL_FePSe3MagExp_DOI}.
  As access to both portals requires user registration, the raw
  diffraction data are also available
\replyv{via the open-access data repository: \href{https://doi.org/10.25500/edata.bham.00001814}{https://doi.org/10.25500/edata.bham.00001814}},
\replyvs{from the corresponding authors upon request}
as are the raw resistivity data, and the
processed data shown in the figures.
Data analysis is detailed in the Methods and Supplementary Information.
}


\section{Acknowledgments}

We thank the Diamond Light Source (DLS), UK for supporting the
synchrotron X-ray experiments through the approved beamtime proposals
CY31687-1 on I19-2, and CY31687-2 on I15.  We thank the Institut
Laue-Langevin (ILL), France for supporting the neutron scattering
experiment through the approved proposal 5-31-2847.  We also thank the
Cubic Anvil Cell (CAC) Station of the Synergetic Extreme Condition
User Facility, China for the support of conducting resistivity
measurements.  We would also like to acknowledge meaningful
discussions with Emilio Artacho, Patricia L. Alireza, Corentin Morice,
Jasper van Wezel, Peter Littlewood and Gil Lonzarich,
and repeated resistivity measurements at CAC by Qingxin Dong.

\section{Funding Statement}
This project has received funding from the
U.K. Department of Science, Innovation and Technologies (DSIT), Grants
No.G115693, to support collaboration between the Cavendish Laboratory
and the Navoi State University of Mining and Technologies.  SD
acknowledges Cambridge Trust, WP Napier PhD Studentship, Cambridge
Philosophical Society Research Studentships for pursuing doctoral
study and the ILL Theory group support for continuing the work.  This
work was supported by a UKRI Future Leaders Fellowship (MJC), Grant
No. MR/Y016602/1.  This work was performed in part at the Aspen Center
for Physics, which is supported by US National Science Foundation
grant PHY-2210452.  The work at SNU (CK, JGP) is funded by the Leading
Researcher Program of the National Research Foundation of Korea (Grant
No. 2020R1A3B2079375).  PY, BW and JC are supported by the National
Key Research and Development Program of China (2023YFA1406100,
2023YFA1607400), National Natural Science Foundation of China
(12025408, 11921004, U23A6003, 12474055, 12404067) and the outstanding
member of Youth Promotion Association of CAS (Y2022004).

\section{Author Contributions Statement}
SD, MJC, CRSH and SSS conceived the project.  SD, MJC, CRSH, HH, XZ,
DD, MRW performed the powder and single-crystal synchrotron
diffraction experiments at DLS.  MJC, DMJ, ARW, TCH, SK performed
powder neutron diffraction at ILL. PY, BW, JC performed resistivity
measurement at CAC.  CK, JGP synthesized the samples.  SD, MJC, CRSH,
HH, GIL and ARW analyzed the data.  DMJ, XZ and CL assisted with the
discussions.  SD, MJC, ARW and SSS wrote the first draft, commented by
CRSH, HH, GIL, DD and SK.  All authors have read and approved
the final manuscript.

\section{Competing Interests Statement}
\replyv{The authors declare no competing interests.}
\replyvs{``There are no competing interests to declare.''}

\section{Additional information}
Supplementary Information: Fig. S1 to S17 and Tables S1 to S7.

Correspondence and requests for materials should be addressed to S.D., M.J.C. or S.S.S.

\end{document}